\documentclass[final,1p,times,twocolumn]{elsarticle}

\usepackage{amsmath}
\usepackage{color}
\usepackage{float}
\usepackage{amsmath}
\usepackage{array}
\usepackage{rotating}
\usepackage{algpseudocode}
\usepackage{algorithm}
\usepackage{subfigure}
\usepackage{soul}
\usepackage{longtable}
\usepackage[normalem]{ulem}
\usepackage[colorlinks=true,linkcolor=blue,citecolor=blue,urlcolor=blue]{hyperref}
\usepackage{multirow}

\newcommand{\bra}[1] { \langle #1 | }
\newcommand{\ket}[1] { | #1 \rangle }
\newcommand{\braket}[2] { \langle #1 | #2 \rangle }
\newcommand{\Bra}[1] { \Bigl\langle #1 \Bigr| }
\newcommand{\Ket}[1] { \Bigl| #1 \Bigr\rangle }

\newcommand{\nks}[0] { N_{\mathbf{k}} }

\newcommand{\nksinv}[0] { \frac{1}{N_{\mathbf{k}}} }

\newcommand{\nqs}[0] { N_{\mathbf{q}} }
\newcommand{\nqsinv}[0] { \frac{1}{N_{\mathbf{q}}} }

\newcommand{\nbnd}[0] { N_{\mathrm{occ}} }

\def\QE{\textsc{Quantum ESPRESSO}\,}
\def\PW{\texttt{PW}\,}
\def\HP{\texttt{HP}\,}

\newcommand{\editor}[2]{%
  \expandafter\newcommand\csname #1note\endcsname[1]{%
    \textcolor{#2}{(\textbf{#1:} ##1)}}%
  \expandafter\newcommand\csname #1\endcsname[1]{%
    \textcolor{#2}{##1}}%
  \expandafter\newcommand\csname #1cancel\endcsname[1]{%
    \textcolor{#2}{\sout{##1}}}%
  \expandafter\newcommand\csname #1change\endcsname[2]{%
    \textcolor{#2}{\sout{##1} ##2}}%
  \newenvironment{#1text}{\color{#2}}{\color{black}}
}

\editor{R}{red}

\begin{document}

\begin{frontmatter}

\title{\texttt{HP} - A code for the calculation of Hubbard parameters using density-functional perturbation theory} 
\author[EPFL]{Iurii Timrov\corref{author}}
\author[EPFL,PSI]{Nicola Marzari}
\author[UniPavia]{Matteo Cococcioni}

\cortext[author] {Corresponding author. \textit{e-mail address:} iurii.timrov@epfl.ch}
\address[EPFL]{Theory and Simulation of Materials (THEOS), and National Centre for Computational Design and Discovery of Novel Materials (MARVEL), \'Ecole Polytechnique F\'ed\'erale de Lausanne (EPFL), CH-1015 Lausanne, Switzerland}
\address[PSI]{Laboratory for Materials Simulations, Paul Scherrer Institut, 5232 Villigen PSI, Switzerland}
\address[UniPavia]{Department of Physics, University of Pavia, via Bassi 6, I-27100 Pavia, Italy} 


\begin{abstract}
We introduce \texttt{HP}, an implementation of density-functional perturbation theory, designed to compute Hubbard parameters (on-site $U$ and inter-site $V$) in the framework of DFT+$U$ and DFT+$U$+$V$. The code does not require the use of computationally expensive supercells of the traditional linear-response approach; instead, unit cells are used with monochromatic perturbations that significantly reduce the computational cost of determining Hubbard parameters. \texttt{HP} is an open-source software distributed under the terms of the GPL as a component of \QE. As with other components, \texttt{HP} is optimized to run on a variety of different platforms, from laptops to massively parallel architectures, using native mathematical libraries (LAPACK and FFTW) and a hierarchy of custom parallelization layers built on top of MPI. The effectiveness of the code is showcased by computing Hubbard parameters self-consistently for the phospho-olivine Li$_x$Mn$_{1/2}$Fe$_{1/2}$PO$_4$ ($x=0, 1/2, 1$) and by highlighting the accuracy of predictions of the geometry and Li intercalation voltages.
\end{abstract}

\end{frontmatter}
{\bf PROGRAM SUMMARY}

\begin{small}
  \noindent
  {\em Program Title:} \texttt{HP} \\
  {\em Licensing provisions:} GNU General Public License V 2.0  \\
  {\em Programming language:} Fortran 95                         \\
  {\em External routines:} \texttt{HP} is a tightly integrated component of the
  \QE\, distribution and requires the standard libraries linked by it: BLAS, LAPACK, FFTW, MPI.\\
  {\em Nature of problem:} Calculation of Hubbard interaction parameters for DFT+$U$ and DFT+$U$+$V$.
  \\
  {\em Solution method:} Hubbard parameters are expressed in terms of the inverse response matrices to localized perturbations of the atomic occupations. The response matrices are computed using density-functional perturbation theory to first order (linear-response theory) in the reciprocal space, that allows to reconstruct the response to a localized perturbation (obtained from calculations in an appropriately sized supercell) as the superposition of the responses to a series of monochromatic perturbations in a primitive unit cell, thus reducing significantly the computational cost. The response matrices are computed via a self-consistent solution of the static Sternheimer equation, whose implementation does not require the calculation of any virtual states. Pseudopotentials (norm-conserving, ultrasoft, projector augmented wave) are used in conjunction with plane-wave basis sets and periodic boundary conditions. 
  \\
  {\em Additional comments including restrictions and unusual features:} Local and semi-local exchange-correlation kernels only. Noncollinear spin-polarized formalism is not supported, only collinear spin-polarized or non-spin-polarized cases can be treated. Spin-orbit coupling cannot be used. Calculation of Hund's $J$ is not supported. Multiple Hubbard channels per atom are not supported. The Hubbard manifold can be only constructed on atomic orbitals, both orthogonalized and non-orthogonalized, while Wannier functions (as well as other localized basis sets) are not supported. The LR approach we adopt here typically results in Hubbard parameters that are unphysically large for closed shell states~\cite{Yu:2014}. No virtual orbitals are used, nor even calculated. \\
  The distribution file of this program can be downloaded from the \QE\, website: http://www.quantum-espresso.org/, and the development version of this program can be downloaded via Git from the GitLab website: https://gitlab.com/QEF/q-e. Interactions with end users of the \HP code happen via a mailing-list forum of \QE: https://www.quantum-espresso.org/forum. Documentation of the \HP code is tightly coupled with the code and is done via standard code comments; different subroutines that implement different equations of the DFPT formalism contain references to the two main papers~\cite{Timrov:2018, Timrov:2021} describing in detail theory behind the implementation. \\
\end{small}

\section{Introduction}
\label{sec:intro}

The development of density-functional theory (DFT)~\cite{Hohenberg:1964, Kohn:1965} 
has allowed modeling of a broad spectrum of properties for a large variety of systems. In practical applications DFT relies on approximations to the exchange-correlation (xc) electronic 
interactions, among which the local-density approximation (LDA) and the generalized-gradient approximation (GGA) are the most popular ones. Both approximations suffer from self-interaction errors (SIE)~\cite{Perdew:1981}, which limit the accuracy only to simple systems with chemistries determined by $s$- and $p$-type electrons. In systems with strongly localized electrons of $d$ and $f$ type, the SIE of LDA and GGA is quite large and leads to a significant overdelocalization of these electrons which translates into quantitative and sometimes even qualitative failures in the description of complex materials.

A popular approach to alleviate SIE in DFT calculations is to use Hubbard corrections to approximate DFT energy functionals ~\cite{anisimov:1991, anisimov:1997, dudarev:1998}. The rationale for this is that Hubbard corrections impose piecewise linearity in the energy functional as a function of atomic occupations~\cite{Cococcioni:2005}, and thus remove/alleviate SIE in the Hubbard manifold both in extended systems and in molecular ones~\cite{Kulik:2006, Kulik:2008}. Within this approach, often referred to as DFT+$U$, the Hubbard correction acts selectively on strongly localized manifolds (of $d$ or $f$ types, typically) through projectors on the corresponding states, while electrons on more delocalized states are treated at the level of approximate DFT. What has made this approach popular is certainly the possibility to achieve significant improvement in the description of systems with localized electrons, while maintaining a combination of simplicity and reduced computational costs. An extended formulation of DFT+$U$ -- so-called DFT+$U$+$V$~\cite{Campo:2010} -- takes into account inter-site Hubbard interactions through the parameter $V$ and allows to improve the description of localized electrons even in presence of significant hybridization with neighbors. Recent works~\cite{Cococcioni:2019, Ricca:2020, Timrov:2021} have highlighted the quantitative accuracy of DFT+$U$+$V$ calculations with both on-site $U$ and inter-site $V$ effective parameters. It is useful to mention about alternative recent formulations of DFT+$U$+$V$~\cite{TancogneDejean:2020, Lee:2020}.

The effectiveness of the (extended) Hubbard functional depends critically on the values of the effective interaction parameters (i.e., the Hubbard $U$ and $V$). Unfortunately, the values of these parameters are not known {\it a priori} and it is still quite a common practice in literature to evaluate them semi-empirically by fitting various experimental properties (when available), which prevents this method from being fully {\it ab initio} and from being predictive for novel materials. Most importantly, it is often forgotten that the Hubbard correction acts on a specific manifold that can be defined in different ways~\cite{Timrov:2020b}. Hence, the values of the Hubbard parameters are not transferable and should not be considered as universal quantities for a given element or material (see the appendix in Ref.~\cite{Kulik:2008}). It is crucial to use the $U$ and $V$ parameters consistently, i.e., maintaining the same Hubbard projectors, pseudopotentials, oxidation states, functionals, etc. for which they were computed. During the past 30 years there has been a large effort to develop methods for the first-principles calculation of Hubbard parameters. Among these, the constrained DFT (cDFT) approach~\cite{Dederichs:1984, Mcmahan:1988, Gunnarsson:1989, Hybertsen:1989, Gunnarsson:1990, Pickett:1998, Cococcioni:2005, Solovyev:2005, Nakamura:2006, Shishkin:2016}, the Hartree-Fock-based approaches~\cite{Mosey:2007, Mosey:2008, Andriotis:2010, Agapito:2015, TancogneDejean:2020, Lee:2020}, and the constrained random phase approximation (cRPA) approach~\cite{Springer:1998, Kotani:2000, Aryasetiawan:2004, Aryasetiawan:2006, Sasioglu:2011, Vaugier:2012, Amadon:2014} are the most popular. A linear-response formulation of cDFT (LR-cDFT) was introduced in Ref.~\cite{Cococcioni:2005} and generalized to the calculation of the inter-site Hubbard parameters $V$ in Ref.~\cite{Campo:2010} (see also Refs.~\cite{Kulik:2011, Himmetoglu:2014}). LR-cDFT has recently been recast via density-functional perturbation theory (DFPT)~\cite{Timrov:2018, Timrov:2021}, allowing to overcome several challenges of the supercell approach of Ref.~\cite{Cococcioni:2005}. It was shown that both LR-cDFT and DFPT give \textit{exactly the same} values of Hubbard parameters (modulo the numerical noise), as they ought to by construction~\cite{Timrov:2018, Timrov:2021}. By constructing the localized perturbation (typically requiring costly calculations in supercells) as a series of independent monochromatic perturbations in the primitive unit cell, it improves significantly the computational efficiency, accuracy, user-friendliness, and automation~\cite{Timrov:2018, Timrov:2021}, as also demonstrated by several recent applications~\cite{Cococcioni:2019, Ricca:2019, Ricca:2020, Floris:2020, Timrov:2020, KirchnerHall:2021, Xiong:2021, Zhou:2021, Mahajan:2021, Timrov:2022, Mahajan:2022}. Key to this successful implementation of the LR-cDFT is indeed the capability to express perturbation theory in reciprocal space as in the calculation of phonons using DFPT~\cite{Baroni:1987, Giannozzi:1991, Baroni:2001}.

An open and recurring question concerns the extent at which the Hubbard parameters obtained from different approaches, e.g. cRPA~\cite{Aryasetiawan:2004} vs. LR-cDFT/DFPT~\cite{Cococcioni:2005, Timrov:2018} vs. ACBN0~\cite{Agapito:2015} compare to one another. First and foremost, these methods have quite different definitions of the Hubbard $U$ which complicates very much their comparison. In fact, only a few attempts have been made in the literature trying to identify the analogies and differences between these theories~\cite{Yu:2014,Aryasetiawan:2006, Himmetoglu:2014}. Moreover, different Hubbard projector functions can be used in these methods: cRPA is based on maximally localized Wannier functions~\cite{Marzari:1999}, while LR-cDFT/DFPT and ACBN0 are used with several other popular types of Hubbard projectors (e.g. nonorthogonalized or orthogonalized atomic orbitals, projector-augmented-wave (PAW) functions, to name a few). Since the computed Hubbard parameters are very sensitive to the choice of Hubbard projector functions~\cite{Mahajan:2021}, the comparison between the values of $U$ computed using different methods and different projector functions becomes even more complicated. Furthermore, in some works the Hubbard $U$ is computed in a one-shot fashion (i.e. a single calculation from an uncorrected ground state) while in other works a self-consistent procedure is adopted where the effective Hubbard $U$ is recomputed from a DFT+$U$ ground state until self-consistency is reached~\cite{Timrov:2021, Cococcioni:2005, Hsu:2009}. The choice of pseudopotentials is also very important when computing the Hubbard parameters~\cite{Kulik:2008, Shishkin:2016}: the value of $U$ obtained for the same structure but with differing pseudopotentials may easily differ by as much as $2 – 3$ eV, in particular if the pseudopotentials were generated from different oxidation states~\cite{Kulik:2008}. All in all, for a given material of interest, even if the calculations of $U$ are done using the same crystal structure and atomic positions, the same kinetic-energy cutoff, $\mathbf{k}$ points sampling of the Brillouin zone (BZ) and same other technical details, all other aforementioned aspects must be very carefully considered when comparing studies reporting different Hubbard $U$ parameters. Therefore, we believe that it makes sense to compare $U$ values computed using different methods only when the same computational setup is used.

In this paper we introduce a computer code, named \HP (\textbf{H}ubbard \textbf{P}arameters), which implements DFPT for the calculation of Hubbard parameters. \HP is distributed under the terms of the GPL license~\cite{GPL}, as a component of the \QE suite of open-source codes based on plane-wave basis sets, pseudopotentials, and using periodic boundary conditions \cite{Giannozzi:2009, Giannozzi:2017, Giannozzi:2020}. 
 
This paper is organized as follows. In Sec.~\ref{sec:theory} we provide a theoretical background for the Hubbard-corrected DFT and for DFPT. In Sec.~\ref{sec:Description_of_software_components} we describe the components of \HP. In Sec.~\ref{sec:installation_and_parallelization} we provide the instructions for installing \HP on UNIX systems and discuss various levels of parallelization implemented in it. In Sec.~\ref{sec:benchmarking} we give an example of the usage of \HP for the calculation of Hubbard parameters for Li$_x$Mn$_{1/2}$Fe$_{1/2}$PO$_4$ and studying its ground-state properties. Finally, Sec.~\ref{sec:Conclusions} contains conclusions and perspectives for future work. \ref{sec:Sample_input_files} presents examples of the input files for the \PW and \HP codes, and \ref{sec:Input_variables} contains the description of the input variables of the \HP code. Hartree atomic units are used throughout the paper.

\section{Theory}
\label{sec:theory}

\subsection{Hubbard-corrected density-functional theory}
\label{sec:DFT_Hubbard}

In this section we briefly review the formulation of the extended DFT+$U$+$V$ approach~\cite{Campo:2010}. All equations which are presented below can be easily reduced to the case of DFT+$U$ by simply setting $V=0$. For the sake of simplicity, the formalism is presented in the framework of norm-conserving (NC) pseudopotentials (PPs), for non-metallic ground states, in the collinear spin-polarized case.

As a generalization of DFT+$U$, DFT+$U$+$V$ is also based on an additive correction to the approximate DFT energy functional, modeled on the Hubbard Hamiltonian~\cite{Campo:2010}: 
\begin{equation}
E_{\mathrm{DFT}+U+V} = E_{\mathrm{DFT}} + E_{U+V} ,
\label{eq:Edft_plus_u}
\end{equation}
where $E_{\mathrm{DFT}}$ represents the approximate DFT energy (constructed, e.g., within the local spin density approximation -- LSDA, or the spin-polarized generalized-gradient approximation -- GGA), while $E_{U+V}$ contains the additional Hubbard term. 
At variance with the DFT+$U$ approach, containing only on-site interactions, DFT+$U$+$V$ is based on the extended Hubbard model including also inter-site interactions between an atom and its surrounding ligands. In the simplified rotationally-invariant formulation~\cite{dudarev:1998}, the extended Hubbard term reads:
\begin{equation}
E_{U+V} = \frac{1}{2} \sum_I \sum_{\sigma m_1 m_2} 
U^I \left( \delta_{m_1 m_2} - n^{II \sigma}_{m_1 m_2} \right) n^{II \sigma}_{m_2 m_1}  - \frac{1}{2} \sum_{I} \sum_{J (J \ne I)}^* \sum_{\sigma m_1 m_2} V^{I J} 
n^{I J \sigma}_{m_1 m_2} n^{J I \sigma}_{m_2 m_1} \,,
\label{eq:Edftu}
\end{equation}
where $I$ and $J$ are atomic site indices, $m_1$ and $m_2$ are the magnetic quantum numbers associated with a specific angular momentum, $U^I$ and $V^{I J}$ are the on-site and inter-site Hubbard parameters, and the star in the sum denotes that for each atom $I$, the index $J$ covers all its neighbors up to a given distance (or up to a given shell). 

The atomic occupation matrices $n^{I J \sigma}_{m_1 m_2}$ are based on a generalized projection of the Kohn-Sham (KS) states on localized orbitals $\varphi^{I}_{m_1}(\mathbf{r})$ of neighbor atoms: 
\begin{equation}
n^{I J \sigma}_{m_1 m_2} = \sum_{\mathbf{k}}^{\nks} \sum_v^{\nbnd} 
\bra{\psi^\circ_{v\mathbf{k}\sigma}} \hat{P}^{J I}_{m_2 m_1}
\ket{\psi^\circ_{v\mathbf{k}\sigma}} \,, 
\label{eq:occ_matrix_0}
\end{equation}
where $v$ and $\sigma$ represent, respectively, the band and spin labels of the KS (pseudo-)wavefunctions, $\mathbf{k}$ indicate points (wave vectors) in the first BZ, $\nks$ being their number, $\nbnd$ is the number of occupied KS states, and $\hat{P}^{J I}_{m_2 m_1}$ is the generalized projector on the localized orbitals of neighbor atoms:
\begin{equation}
\hat{P}^{J I}_{m_2 m_1} = 
\ket{\varphi^{J}_{m_2}} \bra{\varphi^{I}_{m_1}} \,.
\label{eq:Pm1m2}
\end{equation}
In Eq.~\eqref{eq:occ_matrix_0} and hereafter, with the superscript $^\circ$ we indicate quantities which refer to the unperturbed ground state of the system. Here, $\varphi^I_{m_1}(\mathbf{r}) \equiv \varphi^{\gamma(I)}_{m_1}(\mathbf{r} - \mathbf{R}_I)$ are localized orbitals centered on the $I^\mathrm{th}$ atom of type $\gamma(I)$ at the position $\mathbf{R}_I$. Given their importance for the calculation of the Hubbard parameters it is convenient to establish a short-hand notation for the on-site terms of the quantities defined in Eqs.~\eqref{eq:occ_matrix_0} and ~\eqref{eq:Pm1m2}: $n^{I\sigma}_{m_1m_2} \equiv n^{II\sigma}_{m_1m_2}$ and $\hat P^I_{m_1 m_2} \equiv \hat P^{II}_{m_1 m_2}$. The standard DFT+$U$ approach is recovered by setting $V^{IJ}=0$ in Eq.~\eqref{eq:Edftu}. Based on the definitions above, it is quite straightforward to see from Eq.~\eqref{eq:Edftu} that the two terms in the corrective energy functional, proportional to the on-site ($U^{I}$) and inter-site ($V^{IJ}$) couplings, respectively, counteract each other. In fact, while the on-site term favors localization on atomic sites (typically suppressing hybridization), the inter-site one favors hybridized states with components on neighbor atoms. Computing the value of the $U^I$ and $V^{IJ}$ effective interaction parameters is thus crucial to determine the degree of atomic localization of $d$- and/or $f$-type electrons when the system is in its ground state. The Hubbard projector functions $\{ \varphi^{I}_{m_1}(\mathbf{r}) \}$ can be constructed using different types of projector functions as a basis set (see Sec.~\ref{sec:Hubbard_projectors}).

The action of the Hubbard potential on KS wavefunctions can be obtained by taking the functional derivative of $E_{\mathrm{DFT}+U+V}$ [see Eq.~\eqref{eq:Edft_plus_u}] with respect to the complex conjugate of the same KS wavefunction~\cite{Campo:2010, Timrov:Note:2018:errorJPCM}. The term corresponding to this functional derivative of $E_{U+V}$ [see Eq.~\eqref{eq:Edftu}] is:
\begin{equation}
\hat{V}^\circ_{\mathrm{Hub},\sigma} = 
\sum_I \sum_{m_1 m_2} U^{I} \left( \frac{\delta_{m_1 m_2}}{2} - 
n^{I \sigma}_{m_1 m_2} \right) \hat{P}^{I}_{m_1 m_2} - \sum_{I} \sum_{J (J \ne I)}^* \sum_{m_1 m_2} V^{I J} n^{I J \sigma}_{m_1 m_2} \hat{P}^{I J}_{m_1 m_2} \,.
\label{eq:Hub_pot_0}
\end{equation}
This Hubbard potential is added to the standard DFT Hamiltonian~\cite{Timrov:2018, Timrov:2021}, and then the modified KS equations are solved self-consistently.

As was mentioned in Sec.~\ref{sec:intro}, in DFT+$U$ and DFT+$U$+$V$ the two crucial aspects are: 
$i)$~the choice of projector functions that are used to construct the Hubbard manifold, and $ii)$~the choice of the Hubbard parameters ($U^I$ and $V^{IJ}$). It is of utmost importance to understand that the latter strongly depends on the former. Therefore, it makes sense to discuss the values of Hubbard parameters \textit{only} when the Hubbard projectors have been fixed. 

The use of DFT+$U$+$V$ instead of DFT+$U$ is mainly motivated by the increased flexibility the former offers compared to the latter. This is particularly important for more covalently-bonded systems where electronic localization occurs on hybridized states with components on neighbor atoms. In fact, a too large $U$, favoring on-site localization, might suppress the inter-site hybridization and distort the electronic structure of these systems to the point that a negative $U$ might be required to restore a more physical picture and to recover the results achieved with hybrid HSE06 functionals~\cite{Yu:2020}. This effect was already observed in Ref.~\cite{Campo:2010} that showed that the standard on-site only DFT+$U$ can actually suppress the insulating behavior of covalent semiconductors (Si and GaAs) and decrease their band gap. In the same work it was first demonstrated that the use of a finite and positive $V$ (within DFT+$U$+$V$), by favoring the hybridization of atomic states between neighbor atoms, can in fact re-establish the insulating character of their ground state also predicting a wider band gap than the one obtained using uncorrected DFT functionals. A finite positive $V$ was also found effective in re-establishing a sound description of other problematic covalent systems without the need of  negative on-site $U$'s, as detailed in Ref.~\cite{Lee:2020}. While the need of an inter-site interaction $V$ makes a semi-empirical evaluation of the Hubbard parameters much harder, the calculation of $V$ from first-principles can be achieved simultaneously to that of $U$, as illustrated already in a number of studies~\cite{Timrov:2018, Timrov:2021, Campo:2010, TancogneDejean:2020, Lee:2020}. Therefore, DFT+$U$+$V$ where both $U$ and $V$ values are computed \textit{ab initio} constitutes a robust and accurate approach that describes accurately the on-site localization and inter-site hybridization of electrons without any manual calibrations of Hubbard parameters. More on this is discussed in Sec.~\ref{sec:DFPT}.

\subsection{Hubbard projectors} 
\label{sec:Hubbard_projectors}

As was discussed above, one of the key aspects of the Hubbard-corrected DFT formalism is the choice of the projector functions for the Hubbard manifold. In other words, we need to choose the basis $\{ \varphi^I_{m}(\mathbf{r}) \}$ for the projector $\hat{P}^{J I}_{m_2 m_1}$ [see Eq.~\eqref{eq:Pm1m2}]. There are quite many possible projector functions to use as a basis for the Hubbard manifold (see e.g. Refs.~\cite{Tablero:2008, Wang:2016}). In particular, we mention here nonorthogonalized~\cite{Cococcioni:2005, Amadon:2008} and orthogonalized atomic orbitals~\cite{Cococcioni:2019, Ricca:2020, Timrov:2020}, nonorthogonalized~\cite{ORegan:2010} and orthogonalized Wannier functions~\cite{Korotin:2012}, linearized augmented plane-wave approaches~\cite{Shick:1999}, and PAW projector functions~\cite{Bengone:2000, Rohrbach:2003}. A common feature of all these projector functions is that they are spatially localized and depend explicitly on atomic positions. In this work, we consider only two types of projector functions, nonorthogonalized and orthogonalized atomic orbitals. Let us comment briefly about each of them. 

Nonorthogonalized atomic orbitals represent one of the simplest choices for the Hubbard projectors, and they are well suited for systems with a pronounced ionic character. These functions are provided with pseudopotentials and, by virtue of their angular part, they are orthonormal within each atom but not between different atoms. Therefore, whenever inter-atomic overlaps between them become relevant, this type of projector functions is not suitable any more, and inter-site orthogonalization becomes necessary. 
Orthogonalized atomic orbitals are obtained by taking atomic orbitals of each atom and then orthogonalizing them to all the orbitals of all the atoms in the system. In this work, we will use the L\"owdin orthogonalization method~\cite{Lowdin:1950, Mayer:2002}. By doing so, a new set of orbitals is obtained that, by virtue of their mutual orthogonalization~\cite{Timrov:2020b}, provide a more accurate representation of inter-site hybridization. This choice is particularly appealing to define the Hubbard projectors, because it allows us to avoid counting Hubbard corrections twice in the interstitial regions between atoms. This is especially relevant in the case of DFT+$U$+$V$. As explained in Ref.~\cite{Timrov:2020b},  L\"owdin orthogonalized atomic orbitals are defined as:
\begin{equation}
    \varphi^I_{m_1}(\mathbf{r}) = \sum_{J m_2} \left(O^{-\frac{1}{2}}\right)^{JI}_{m_2 m_1} \phi^J_{m_2}(\mathbf{r}) \,,
    \label{eq:OAO_def}
\end{equation}
where $O$ is the orbital overlap matrix which is defined from its matrix elements as: $(O)^{IJ}_{m_1 m_2} = \braket{\phi^I_{m_1}}{\phi^J_{m_2}}$, where $\phi^I_{m_1}(\mathbf{r})$ and $\phi^J_{m_2}(\mathbf{r})$ are the nonorthogonalized atomic orbitals. It is important to note that the atomic and state indices must be understood as being in couples, $(I,m_1)$ and $(J,m_2)$, because for different types of atoms the considered atomic manifolds can be different. Here we orthogonalize \textit{all} the states of \textit{all} the atoms in the system. It is important to orthogonalize not only states that belong to the chosen Hubbard manifolds of each atom (e.g., $d$ or $f$ states), but also the remaining states, in order to preserve the on-site orthogonality.

In practice, we construct Bloch sums of the L\"owdin orthogonalized atomic orbitals and then use only their lattice-periodic parts (see Eqs.~(A2) and (A5) in Ref.~\cite{Timrov:2018}). Further, these quantities are Fourier-transformed from real to reciprocal space and used in the DFT+$U$(+$V$) and DFPT formalisms to compute various scalar products [see e.g. Eqs.~\eqref{eq:occ_matrix_0} and \eqref{eq:Pm1m2}] as sums over reciprocal lattice vectors.

As shown in Ref.~\cite{Mahajan:2021}, the computed Hubbard parameters do depend strongly on the choice of the Hubbard projector functions. For example, for $\beta$-MnO$_2$ the $U$ values computed using nonorthogonalized and orthogonalized atomic orbitals differ by about $1-2$~eV while $V$ values differ by about 0.3~eV~\cite{Mahajan:2021}. Therefore, it is crucial that the same Hubbard projector functions are used for computing Hubbard parameters and for the subsequent DFT+$U$(+$V$) production calculations.

\subsection{Hubbard parameters from density-functional perturbation theory} 
\label{sec:DFPT}

Following Ref.~\cite{Cococcioni:2005}, Hubbard parameters can be defined as the second derivatives of the total energy with respect to the total occupation of a given atom, i.e. with respect to the trace of the site-diagonal occupation matrix $n^{I\sigma}_{m_1 m_2}$. This is consistent with the structure of the simple Hubbard corrective functional, shown in Eq.~\eqref{eq:Edftu}, that subtracts from the total energy a quadratic term in the atomic occupation, to substitute it with a linear one, thus removing the unphysical self-interaction (delocalization) errors from approximate energy functionals. In this context, the inter-site part in Eq.~\eqref{eq:Edftu} serves as an off-diagonal correction when removing self-interactions, which is especially relevant for systems with a covalent bonding. In practice, the calculation of the energy second derivative can be achieved by perturbing the system with a shift in the potential acting on the Hubbard states of a given atom $J$, $\lambda^J \sum_m \hat{P}^{J}_{m m}$ ($\lambda^J$ is the strength of the perturbation), and then computing the response of all the atomic occupations. Applying this to all the Hubbard atoms in the system allows to construct the {\it bare} and {\it self-consistent} susceptibility matrices that are obtained, respectively, at the beginning of the perturbed run and at its self-consistent convergence:
\begin{equation}
    (\chi_0)_{IJ} = \frac{dn^I_0}{d\lambda^J}, \qquad
    (\chi)_{IJ} = \frac{dn^I}{d\lambda^J} \,.
    \label{chi}
\end{equation}
Here, $n^I = \sum_{m \sigma} n^{I \sigma}_{m m}$ and similar definition holds for $n^I_0$.
From these, the effective Hubbard parameters can be readily obtained~\cite{Cococcioni:2005}:
\begin{equation}
U^I = \left( \chi_0^{-1} - \chi^{-1} \right)_{II} \,,
\label{eq:U_def}
\end{equation} 
\begin{equation}
V^{IJ} = \left( \chi_0^{-1} - \chi^{-1} \right)_{IJ} \,.
\label{eq:V_def}
\end{equation}
It is important to stress that the procedure outlined above is based on isolated perturbations; therefore, it requires the use of large supercells (whose size has to be increased until the convergence of $U^I$ and $V^{IJ}$ is achieved)~\cite{Cococcioni:2005} that makes these calculations computationally demanding and prone to accuracy issues due to their problematic convergence and their reliance on small energy differences.  

DFPT offers a more efficient approach to linear-response calculations and allows us to largely reduce these issues~\cite{Timrov:2018, Timrov:2021}.
Within the framework of DFPT, the response of the KS wavefunctions to a small perturbation of the atomic potential [that induces a variation of the atomic occupations defined in Eq.~\eqref{chi}] is obtained as the self-consistent solution of the perturbative problem resulting from a first-order variation of the KS equations:
\begin{equation}
\left( \hat{H}^\circ_{\sigma} - \varepsilon^\circ_{v\mathbf{k}\sigma} 
\, \right) \Ket{\frac{d\psi_{v\mathbf{k}\sigma}}{d\lambda^{J}}} = - \biggl( \frac{d\hat{V}_{\mathrm{Hxc},\sigma}}{d\lambda^{J}} 
- \frac{d \varepsilon_{v\mathbf{k}\sigma}}{d\lambda^{J}}
+ \hat{V}_\mathrm{pert}^{J} \biggr) \, \ket{\psi^\circ_{v\mathbf{k}\sigma}} \,,
\label{eq:KS_lin_eq_q}
\end{equation}
where $\hat{H}^\circ_{\sigma} = \hat{H}^\circ_{\mathrm{DFT},\sigma} + \hat{V}^\circ_{\mathrm{Hub},\sigma}$ is the total Hamiltonian of DFT+$U$+$V$, where $\hat{H}^\circ_{\mathrm{DFT},\sigma}$ is the Hamiltonian of standard DFT, and $\hat{V}^\circ_{\mathrm{Hub},\sigma}$ is the corrective Hubbard potential defined in Eq.~\eqref{eq:Hub_pot_0}. $\varepsilon^\circ_{v\mathbf{k}\sigma}$ and $\psi^\circ_{v\mathbf{k}\sigma}$ are the KS energies and wavefunctions of the unperturbed system in the DFT+$U$+$V$ framework. $\hat{V}_\mathrm{pert}^{J} = \sum_m \hat{P}^{J}_{mm}$ is the perturbing potential; $\frac{d\hat{V}_{\mathrm{Hxc},\sigma}}{d\lambda^{J}}$, $\frac{d\psi_{v\mathbf{k}\sigma}}{d\lambda^{J}}$, and $\frac{d \varepsilon_{v\mathbf{k}\sigma}}{d\lambda^{J}}$ are the response Hartree and xc (Hxc) potential, response KS wavefunctions, and response KS energies, respectively~\cite{Timrov:2018, Timrov:2021}. It is important to remark that the response of the Hubbard potential is not present in Eq.~\eqref{eq:KS_lin_eq_q} so that the Hubbard parameters are obtained, consistently with their definition, as second derivatives of the DFT part only of the total energy~\cite{Timrov:2018}. The problem has to be solved self-consistently because the response of the KS eigenvalues and of the Hxc potential appearing on the right-hand side of Eq.~\eqref{eq:KS_lin_eq_q} depend on the response of the KS wavefunctions, obtained from the solution of the perturbative problem in the equation above. Once convergence is achieved, the variation of the diagonal (with respect to atomic sites) atomic occupation matrices [that define the self-consistent susceptibility matrix in Eq.~\eqref{chi}] are obtained~\cite{Timrov:Note:2020:nij}: 
\begin{equation}
\frac{dn^{I \sigma}_{m_1 m_2}}{d\lambda^{J}} =
\sum_{\mathbf{k}}^{\nks} \sum_v^{\nbnd} \,
\biggl[ \Bra{\psi^\circ_{v\mathbf{k}\sigma}} \hat{P}^{I}_{m_2 m_1} 
\Ket{\frac{d{\psi}_{v\mathbf{k}\sigma}}{d\lambda^{J}}} + \, \Bra{\frac{d{\psi}_{v\mathbf{k}\sigma}}{d\lambda^{J}}} 
\hat{P}^{I}_{m_2 m_1} \Ket{\psi^\circ_{v\mathbf{k}\sigma}} \biggr] \,.
\label{eq:occ_matrix_response_new}
\end{equation}
The major advantage offered by the DFPT reformulation of LR-cDFT consists in the possibility to obtain the variation of atomic occupations as a sum of wavevector-specific contributions that can be computed independently from one another (thus leading to better scaling of the computational cost~\cite{Timrov:2018}) using the primitive unit cell of the system. In fact, the Fourier spectrum of a perturbation that has the periodicity of a supercell (as needed to eliminate the interactions with  periodic replicas) only contains fundamental vectors of its reciprocal lattice that map into a corresponding $\mathbf{q}$ points grid within the BZ corresponding to the primitive cell~\cite{Timrov:2018}. The total response of atomic occupations can thus be written as (see Eq.~(42) 
in Ref.~\cite{Timrov:2018}):
\begin{equation}
\frac{dn^{s l \sigma}_{m_1 m_2}}{d\lambda^{s' l'}} =
\nqsinv \sum_\mathbf{q}^{\nqs} e^{i\mathbf{q}\cdot(\mathbf{R}_{l}-\mathbf{R}_{l'})} \, 
\Delta^{s'}_\mathbf{q} \bar{n}^{s \, \sigma}_{m_1 m_2} \,,
\label{eq:occ_matrix_response_lp_1}
\end{equation}
where the atomic site indices $I$ and $J$ have been decomposed as $I = (l,s)$ and $J = (l',s')$ indicating, respectively, the cell the atom belongs to ($l$ and $l'$) and its position within the cell ($s$ and $s'$). Here, $N_\mathbf{q}$ is the number of $\mathbf{q}$ points in the first BZ (note that the dimension of the $\mathbf{q}$ points grid reflects directly that of the supercell it is the reciprocal-space image of). Hereafter, we use the over-bar to indicate lattice-periodic parts of the ground-state and response quantities. $\Delta^{s'}_\mathbf{q} \bar{n}^{s \, \sigma}_{m_1 m_2}$ represents the lattice-periodic response of the occupation matrix to a monochromatic perturbation with a wavevector $\mathbf{q}$, and it can be linked to the lattice-periodic variations of the KS wavefunctions as follows \cite{Timrov:2018}:
\begin{equation}
\Delta^{s'}_\mathbf{q} \bar{n}^{s \, \sigma}_{m_1 m_2} =
\nksinv \sum_{\mathbf{k}}^{\nks} \sum_{v}^{\nbnd}
\biggl[ \Bra{\bar{u}^\circ_{v\mathbf{k}\sigma}} 
\hat{\bar{P}}^{s}_{m_2,m_1,\mathbf{k},\mathbf{k}+\mathbf{q}}
\Ket{\Delta^{s'}_\mathbf{q} \bar{u}_{v\mathbf{k}\sigma}} + \Bra{\bar{u}^\circ_{v\mathbf{k}\sigma}} 
\hat{\bar{P}}^{s}_{m_1,m_2,\mathbf{k},\mathbf{k}+\mathbf{q}}
\Ket{\Delta^{s'}_\mathbf{q} \bar{u}_{v\mathbf{k}\sigma}} \biggr] \,.
\label{eq:occ_matrix_response_lp_2}
\end{equation}
Here, $\bar{u}^\circ_{v\mathbf{k}\sigma}$ and $\Delta^{s'}_\mathbf{q} \bar{u}_{v\mathbf{k}\sigma}$ are the lattice-periodic parts of the unperturbed and linear-response monochromatic $\mathbf{q}$ component of the KS wavefunctions, respectively (see the appendices~1 and 3 of Ref.~\cite{Timrov:2018}). The lattice-periodic part of the projector on the Hubbard manifold, which appears in Eq.~\eqref{eq:occ_matrix_response_lp_2}, reads~\cite{Timrov:2018}:
\begin{equation}
\hat{\bar{P}}^{
s}_{m_2,m_1,\mathbf{k},\mathbf{k+q}} = 
\ket{\bar{\varphi}^{s}_{m_2,\mathbf{k}}}
\bra{\bar{\varphi}^s_{m_1,\mathbf{k+q}}} \,.
\label{eq:P_proj_lp0}
\end{equation}
The two terms on the right-hand side of Eq.~\eqref{eq:occ_matrix_response_lp_2} were made look similar (except for the inversion in the order of indices $m_1$ and $m_2$) by the use of time-reversal symmetry. As was mentioned above, due to the linear character of the perturbative problem [Eq.~\eqref{eq:KS_lin_eq_q}], the lattice-periodic components of the response KS wavefunctions at different $\mathbf{q}$ can be computed independently from one another as solutions of $\mathbf{q}$-specific Sternheimer equations~\cite{Timrov:2018}:
\begin{equation}
\left( \hat{\bar{H}}^\circ
_{\mathbf{k+q},\sigma} + 
\alpha \hat{\bar{\mathcal{O}}}_{\mathbf{k+q},\sigma} 
- \varepsilon^\circ_{v\mathbf{k}\sigma} \right) \, 
\ket{\Delta^{s'}_\mathbf{q} \bar{u}_{v\mathbf{k}\sigma}} = - \hat{\bar{\mathcal{P}}}_{\mathbf{k+q}, \sigma}
\left( \, \Delta^{s'}_\mathbf{q} 
\hat{\bar{V}}_{\mathrm{Hxc},\sigma} +
\hat{\bar{V}}^{s'}_{\mathrm{pert}, \mathbf{k+q}, \mathbf{k}} \right) \, 
\ket{\bar{u}^\circ_{v\mathbf{k}\sigma}} \,, 
\label{eq:LRKSeq_lp}
\end{equation}
where the perturbing potential reads:
\begin{equation}
\hat{\bar{V}}^{s'}_{\mathrm{pert}, \mathbf{k}+\mathbf{q},\mathbf{k}} = 
\sum_m \hat{\bar{P}}^{s'}_{m,m,\mathbf{k}+\mathbf{q},\mathbf{k}}\,.
\label{vpert}
\end{equation}
The quantities $\hat{\bar{H}}^\circ_{\mathbf{k+q},\sigma}$ and $\Delta^{s'}_\mathbf{q} \hat{\bar{V}}_{\mathrm{Hxc},\sigma}$ are, respectively, the lattice-periodic parts of the unperturbed Hamiltonian $\hat{H}^\circ_\sigma$ (which contains the Hubbard corrective potential with on-site $U$ and inter-site $V$) and response Hxc potential for a specific $\mathbf{q}$. The response Hxc potential $\Delta^{s'}_\mathbf{q} \hat{\bar{V}}_{\mathrm{Hxc},\sigma}$ depends on the response spin charge density at the same $\mathbf{q}$, $\Delta^{s'}_\mathbf{q} \bar{\rho}_\sigma(\mathbf{r})$, which in turn depends on $\Delta^{s'}_\mathbf{q} \bar{u}_{v\mathbf{k}\sigma}(\mathbf{r})$ (see Ref.~\cite{Timrov:2018} for more details). The operators $\hat{\bar{\mathcal{O}}}_{\mathbf{k+q},\sigma}$ and $\hat{\bar{\mathcal{P}}}_{\mathbf{k+q}, \sigma}$
are the lattice-periodic parts of projectors on the valence and conduction manifolds, respectively~\cite{Baroni:2001,Dalcorso:2001}:
\begin{equation}
\hat{\bar{\mathcal{O}}}_{\mathbf{k+q},\sigma} =  \sum_{v'}^{\nbnd} 
\ket{\bar{u}^\circ_{v'\mathbf{k+q}\sigma}} \bra{\bar{u}^\circ_{v'\mathbf{k+q}\sigma}} \,,
\label{eq:proj_v}
\end{equation}
and
\begin{equation}
\hat{\bar{\mathcal{P}}}_{\mathbf{k+q},\sigma} = 
1 - \sum_{v'}^{\nbnd} \ket{\bar{u}^\circ_{v'\mathbf{k+q}\sigma}} \bra{\bar{u}^\circ_{v'\mathbf{k+q}\sigma}} \,.
\label{eq:proj_c}
\end{equation}
In Eq.~\eqref{eq:LRKSeq_lp}, $\alpha = 2 \, (\mathrm{max}[\varepsilon^\circ_{v\mathbf{k}\sigma}] - \mathrm{min}[\varepsilon^\circ_{v\mathbf{k}\sigma}])$, where $\mathrm{max}[\varepsilon^\circ_{v\mathbf{k}\sigma}]$ and
$\mathrm{min}[\varepsilon^\circ_{v\mathbf{k}\sigma}]$ are the highest and the lowest energies of the occupied KS bands, respectively. The operator $\hat{\bar{\mathcal{O}}}_{\mathbf{k+q},\sigma}$ is inserted on the left-hand side of Eq.~\eqref{eq:LRKSeq_lp} in order to avoid singularity issues during the iterative solution; at the same time the operator $\hat{\bar{\mathcal{P}}}_{\mathbf{k+q},\sigma}$ avoids very expensive sums over numerous empty states~\cite{Baroni:2001, Timrov:2018, Timrov:2021}. Note that due to the presence of the projector $\hat{\bar{\mathcal{P}}}_{\mathbf{k+q},\sigma}$ in Eq.~\eqref{eq:LRKSeq_lp} the derivative of the KS eigenvalues disappears from the right-hand side of Eq.~\eqref{eq:LRKSeq_lp} in comparison to Eq.~\eqref{eq:KS_lin_eq_q} (see also Ref.~\cite{Timrov:2018}). The KS wavefunctions at $\mathbf{k+q}$ points, $\bar{u}^\circ_{v'\mathbf{k+q}\sigma}(\mathbf{r})$, which are present in Eqs.~\eqref{eq:proj_v} and \eqref{eq:proj_c} are obtained by solving non-self-consistently the KS equations and using the unperturbed ground-state spin charge density $\bar{\rho}^\circ_\sigma(\mathbf{r})$.
All the operators in Eq.~\eqref{eq:LRKSeq_lp} appear with a specific $\mathbf{q}$ component as a result of recasting Eq.~\eqref{eq:KS_lin_eq_q} in reciprocal space through the Bloch sums of all the quantities appearing in there (this is discussed in detail in Ref.~\cite{Timrov:2018}). The potential terms appearing on the right-hand side of Eq.~\eqref{eq:LRKSeq_lp} represent the lattice-periodic components of the corresponding potential variations at the indicated wavevector~$\mathbf{q}$. Once these equations are solved self-consistently for all the wavevectors, Eqs.~\eqref{eq:occ_matrix_response_lp_1} and \eqref{eq:occ_matrix_response_lp_2} are used to compute the susceptibility matrices using Eq.~\eqref{chi}, from which the Hubbard interaction parameters are readily obtained as indicated in Eqs.~\eqref{eq:U_def} and \eqref{eq:V_def}.

\subsection{Extensions of the DFPT formalism}

The formalism presented above has been generalized in several ways: $\emph{i)}$ metallic ground states; $\emph{ii)}$ ultrasoft (US) PPs~\cite{Vanderbilt:1990} and the PAW method~\cite{Blochl:1994}; and $\emph{iii)}$ explicit account of symmetry. A detailed discussion about the first two points can be found in Ref.~\cite{Timrov:2021}. 

\textit{Metallic ground states.} The ground state of a given system computed using standard DFT can be used as a starting point for the DFPT calculation of Hubbard parameters. However, in some systems standard DFT predicts a metallic ground state due to large self-interactions errors, while in reality the system is insulating. Therefore, the DFPT approach presented above must be generalized to be able to work on top of metallic ground states. For this reason, an extension of the DFPT approach to metallic ground states has been developed and discussed in detail in Ref.~\cite{Timrov:2021}. It is based on the use of the smearing technique~\cite{deGironcoli:1995}. In metals, very dense $\mathbf{k}$ points sampling is needed to sample the Fermi surface, which is computationally very expensive. Thanks to the smearing of the Fermi surface it is possible to reduce greatly the number of $\mathbf{k}$ points needed to describe electronic states around the Fermi level, which helps containing the computational cost; as a consequence, these states have partial occupancy (between 0 and 1). The DFPT approach can still have the same form as for non-metallic systems but the following modifications are required~\cite{Timrov:2021}: $\emph{i)}$~electronic states should be allowed to have a fractional occupancy which translates into weighted sums over $\mathbf{k}$ points and band indices, $\emph{ii)}$~the definition of the projector $\hat{\bar{\mathcal{P}}}_{\mathbf{k+q},\sigma}$ is generalized, and $\emph{iii)}$~the response occupation matrix $\Delta^{s'}_\mathbf{q} \bar{n}^{s \, \sigma}_{m_1 m_2}$ and the response spin charge density acquire an extra term proportional to the shift of the Fermi level when $\mathbf{q} = \mathbf{0}$. The interested reader can find a detailed discussion in Ref.~\cite{Timrov:2021}.

\textit{US PPs and PAW.} In systems containing localized valence states (e.g., semicore states included in the valence region or atomic states of $d$ or $f$ kind in transition-metal and rare-earth compounds), high kinetic energy cut-offs in the plane-wave expansion are needed if NC PPs are used. In this case, it is useful to replace NC PPs by US PPs or PAW which allow to reduce significantly the kinetic energy cut-off and thus lower the overall computational cost of the calculations. However, when US PPs or PAW are used augmentation terms are added in the expressions for the spin charge density to restore the correct normalization, and consequently extra terms appear in the expressions for the local and Hxc potentials. In US and PAW formalisms, the standard KS equations must be replaced by generalized KS equations which contain the overlap operator $\hat{S}$~\cite{Vanderbilt:1990}, and therefore also the first-order response KS equations are modified~\cite{Timrov:2021}. The unperturbed and response occupation matrices are also generalized, as well as the Hubbard potential, with the projector on the Hubbard manifold acquiring the $\hat{S}$ operator, as $\hat{P}^{J I}_{m_2 m_1} = \hat{S} \ket{ \varphi^{J}_{m_2}} \bra{\varphi^{I}_{m_1}} \hat{S}$. Detailed discussions are presented in Ref.~\cite{Timrov:2021}.

\textit{Symmetry.} The CPU time and memory requirements of the DFPT calculation can be significantly reduced by exploiting the symmetry of the system. In fact, as explained in abundant literature (and specifically in Appendix~A.4 of Ref.~\cite{Giannozzi:2009} for \QE) the use of symmetry allows to focus ground-state calculations only on a small portion of the regular $\mathbf{k}$ point grid used to sample the BZ (the so-called irreducible wedge of the BZ (IBZ)). Within DFPT, the use of symmetry is slightly more articulated due to the presence of perturbations. The response of a system (and the derivative of relevant quantities) is typically reconstructed from the response to a series of monochromatic perturbations that are computed one by one. Since a finite wavelength perturbation ($\mathbf{q} \neq 0$) lowers the symmetry of the crystal, at each specific $\mathbf{q}$ point all the ground-state quantities that are needed in DFPT calculations (e.g., KS wavefunctions, Hamiltonian, and eigenvalues) must be recomputed on an extended IBZ that is determined according to the so-called small group of $\mathbf{q}$ (i.e. the group of symmetries such that $\mathcal{S}_\mathbf{q} \, \mathbf{q} = \pm \mathbf{q} + \mathbf{G}$, where $\mathbf{G}$ is a reciprocal lattice vector). The same procedure needs to be repeated for each $\mathbf{q}$ point. Once DFPT calculations are completed for all the inequivalent points of the $\mathbf{q}$-grid, and the response on the other $\mathbf{q}$-points is reconstructed by symmetry, the total response is computed according to Eq.~\eqref{eq:occ_matrix_response_lp_1}. Symmetry is thus important to reduce the workload of both DFT and DFPT calculations. The use of primitive unit cells, and the possibility to reduce the number of $\mathbf{q}$-specific linear-response calculations thanks to symmetry contribute to make these calculations substantially faster than supercell-based LR-cDFT which cannot escape the cubic scaling of DFT calculations with respect to the number of atoms.

\section{Description of software components}
\label{sec:Description_of_software_components}

The \HP code is contained in a module of the \QE distribution~\cite{Giannozzi:2009, Giannozzi:2017, Giannozzi:2020}, and it resides in a self-contained directory HP under the root directory of the \QE tree. The \HP code is tightly integrated in \QE, and it uses many routines from other modules, namely \PW, \texttt{Modules}, and \texttt{LR\_Modules} (see Ref.~\cite{Giannozzi:2020} for more details). In addition, \HP uses various domain-specific mathematical libraries of \QE, such as \texttt{LAXlib} (containing routines to perform linear-algebra operations) and \texttt{FFTXlib} (containing routines to perform Fast Fourrier Transforms (FFT's)). Basic linear-algebra operations (e.g. matrix-matrix and matrix-vector multiplications, scalar products, matrix inversions, etc.) are efficiently performed using BLAS and LAPACK libraries. In the following we discuss the workflow of the calculation of Hubbard parameters.

\subsection{Ground-state calculation}   

In order to compute the Hubbard parameters for a given system, a standard ground-state DFT [or DFT+$U$(+$V$) with some initial guess of $U$ (and $V$)] calculation has to be performed first, yielding the lattice-periodic parts of the unperturbed KS wavefunctions $\bar{u}^\circ_{v\mathbf{k}\sigma}(\mathbf{r})$ and the KS energies
$\varepsilon^\circ_{v\mathbf{k}\sigma}$ for all the occupied states, the ground-state spin charge density $\bar{\rho}^\circ_\sigma(\mathbf{r})$, and the occupation matrix $n^{I\sigma}_{m_1 m_2}$. The information thus obtained is then used as input for the linear-response calculation with the \HP code. This ground-state calculation is performed using the \PW code (\texttt{pw.x} executable), which is one of the key components of the \QE distribution. In \ref{sec:Sample_input_files} sample input files for \texttt{pw.x} are shown (see the input samples 1 and 2). After the successful completion of the ground-state calculation, the \texttt{pw.x} code writes the ground-state quantities mentioned above to disk, together with all the relevant information about the system, including unit cell and atomic positions, pseudopotentials, energy cutoffs, $\mathbf{k}$ point grids, etc. This data is used by the \HP code which reads it from file. Therefore, it is not necessary to redefine the system under study in the input file of \texttt{hp.x}.

\subsection{Linear-response calculation}   

The linear-response calculation of Hubbard parameters ($U$ in the DFT+$U$ framework, or $U$ and $V$ in the DFT+$U$+$V$ framework) is done using the \HP code (\texttt{hp.x} executable). A list of all input variables of \texttt{hp.x} is given in Table~\ref{tab:Table_input_hp.x} of \ref{sec:Input_variables}, and a sample input file for \texttt{hp.x} is given in \ref{sec:Sample_input_files} (see the input sample 3). The total number of linear-response calculations that have to be performed is $N_\mathrm{pert} N_\mathbf{q}$, where $N_\mathrm{pert}$ is the number of Hubbard atoms in the primitive unit cell that have to be perturbed, and $N_\mathbf{q}$ is the number of points in the $\mathbf{q}$ grid.

The size of the response matrices $\chi$ and $\chi_0$ is $N^\mathrm{sc}_\mathrm{H} \times N^\mathrm{sc}_\mathrm{H}$, where $N^\mathrm{sc}_\mathrm{H} = N_\mathrm{H} N_\mathbf{q}$, and $N_\mathrm{H}$ is the total number of Hubbard atoms in the primitive unit cell~\cite{Timrov:2018}. Each column of the response matrices corresponds to the perturbation of a specific Hubbard atom of the primitive unit cell with a specific $\mathbf{q}$. It is possible to reduce the number of linear-response calculations by perturbing only inequivalent Hubbard atoms in the primitive unit cell. The \HP code contains the implementation of several algorithms which find inequivalent Hubbard atoms (i.e. $N_\mathrm{pert}$); this is controlled by the keyword \texttt{find\_atpert} which is described in Table~\ref{tab:Table_input_hp.x}. The default algorithm for finding inequivalent (i.e., to be perturbed separately) Hubbard atoms is based on the comparison of the traces of unperturbed atomic occupation matrices, $\mathrm{Tr}[n^{I\sigma}_{m_1 m_2}]$, which represent the L\"owdin estimate of the number of electrons residing on the Hubbard manifold of a given Hubbard atom $I$. The calculation of the response to the perturbation of inequivalent Hubbard atoms are independent from each other, and hence it is possible to efficiently parallelize these calculations by performing them on different nodes/machines (see Sec.~\ref{sec:parallelization}).

The linear-response calculation for each perturbed Hubbard atom requires solving $N_\mathbf{q}$ independent $\mathbf{q}$-specific Sternheimer equations~\eqref{eq:LRKSeq_lp}. $N_\mathbf{q}$ is 
the number of points in the $\mathbf{q}$ point grid, $N_\mathbf{q}$ = \texttt{nq1} $\times$ \texttt{nq2} $\times$ \texttt{nq3}. The strength of DFPT (see Sec.~\ref{sec:DFPT}) resides in the fact that linear-response calculations at each $\mathbf{q}$ point are independent from other $\mathbf{q}$ points (in linear regime there is no coupling between perturbations at different wavelengths), and hence it is possible to parallelize calculations over $\mathbf{q}$ points as well (see Sec.~\ref{sec:parallelization}). When the self-consistency of each $\mathbf{q}$-specific Sternheimer equation has been reached, the $\mathbf{q}$-specific response occupations matrices $\Delta^{s'}_\mathbf{q} \bar{n}^{s \, \sigma}_{m_1 m_2}$ are computed using Eq.~\eqref{eq:occ_matrix_response_lp_2}. Then, all the responses are summed up using Eq.~\eqref{eq:occ_matrix_response_lp_1}, thus giving one column of the self-consistent response matrix $\chi$ (see Eq.~\eqref{chi}). One column of the bare response matrix $\chi_0$ is computed in a similar way, however the sum of responses using Eq.~\eqref{eq:occ_matrix_response_lp_1} is computed after the first iteration in the self-consistent cycle of Eq.~\eqref{eq:LRKSeq_lp}, i.e. before that the response of the Hxc potential (i.e. $\Delta^{s'}_\mathbf{q} \hat{\bar{V}}_{\mathrm{Hxc},\sigma}$) builds up. Other columns of the response matrices $\chi$ and $\chi_0$ are obtained from the perturbation of other inequivalent Hubbard atoms.

The Sternheimer equations~\eqref{eq:LRKSeq_lp} are solved iteratively and self-consistently using the conjugate-gradient method~\cite{Baroni:2001, Giannozzi:2009}, and using the standard linear-response machinery of \QE~\cite{Giannozzi:2017, Giannozzi:2020}. In order to speed up the convergence of the iterative solution, the scheme of Ref.~\cite{Johnson:1988} is used for mixing the response Hxc potential $\Delta^{s'}_\mathbf{q} \hat{\bar{V}}_{\mathrm{Hxc},\sigma}$~\cite{Timrov:2018}. Table~\ref{tab:Table_input_hp.x} describes the parameters that control the convergence of the Sternheimer equations and of the response matrices $\chi$ and $\chi_0$.

The final step of the \HP calculation is the postprocessing calculation of the Hubbard parameters using Eqs.~\eqref{eq:U_def} and \eqref{eq:V_def}. This step is computationally inexpensive (negligible compared to the linear-response calculation). In this final phase, the missing columns of the response matrices $\chi$ and $\chi_0$ are reconstructed from the available data (i.e. columns which were computed explicitly, as discussed above) exploiting the symmetry of the system. This is done by comparing the inter-atomic distances, atomic types, and the orientation of spin (up or down). Once the full matrices $\chi$ and $\chi_0$ have been reconstructed, they are inverted to compute the interaction matrix as showed in Eqs.~\eqref{eq:U_def} and \eqref{eq:V_def}.

\section{Software installation, parallelization, and testing}
\label{sec:installation_and_parallelization}

\subsection{Installation instructions}
\label{sec:Installation_instructions}

The \HP program is distributed as source code, like the other components of the \QE distribution. Version control is handled using Git and the code is hosted on the GitLab platform~\cite{QuantumESPRESSO:Gitlab}. The installation procedure of \HP is the same as for all other modules of the distribution. \QE (including \HP\!) makes use of GNU autoconf~\cite{GNUautoconf}. The HP repository, which contains the source \HP code, is residing within the \QE tree. The code is compiled with the following commands from within the \QE\,tree:
\begin{equation*}
 \begin{array}{c}
   \texttt{./configure} \\ 
   \texttt{make pw} \\
   \texttt{make hp}
 \end{array}
\end{equation*}
Alternatively, it is possible to use \texttt{cmake}~\cite{cmake} instead of \texttt{./configure}.
Here, the first step sets up the environment (compilers, libraries, etc.), the second step compiles the \PW code (\texttt{pw.x}), and in the third step, the \HP code (\texttt{hp.x}) is compiled. Links to all these executables are created in the \texttt{bin/} directory of the \QE tree. More detailed instructions on the installation can be found in the documentation that comes with the distribution. If problems are encountered during the installation or the use of HP (or of any other code in the \QE distribution) users can also take advantage of the \QE users forum~\cite{QuantumESPRESSO:forum} by posting specific questions about their difficulties. Users intending to make quick tests with HP or to use it for demonstrative or teaching purposes~\cite{Kobayashi:2021} could consider the Quantum Mobile~\cite{QuantumMobile} - a virtual machine that has \QE pre-installed along with several other codes for quantum-mechanical materials simulations. Quantum Mobile can be easily installed on laptops and desktops and its use avoids any issues related to the installation of the \QE package. Of course, the Quantum Mobile is not recommended for production runs; an optimized installation of \QE on workstations and high-performance computers should instead be preferred for these purposes.

\subsection{Parallelization of the code}
\label{sec:parallelization}

Like the other components of \QE\!, the \HP code is optimized to run on a variety of different platforms, from laptops to massively parallel architectures. The parallelization of the
\HP code is achieved by using the message-passing paradigm and calls to standard Message Passing Interface (MPI) libraries~\cite{MPI:1994}. High performance on massively parallel architectures
is achieved by distributing both data and computations in a hierarchical way across processors. 
The FFT's, which are used for transformations from real space to reciprocal space and vice versa, are also efficiently parallelized among processors~\cite{Giannozzi:2009}.
The \HP code supports four levels of parallelization: 
\begin{enumerate}
\item The parallelization over perturbed Hubbard atoms, which is implemented by distributing independent linear-response atom-specific calculations (each having a grid of $\mathbf{q}$ points) across the processors, each taking care of one atom-specific perturbation. This parallelization is controlled by setting the input parameter \texttt{perturb\_only\_atom(i)} to \texttt{.true.}, which specifies that the Hubbard atom with the index \texttt{i} will be perturbed in the current run;
\item The $\mathbf{q}$~points parallelization, which is implemented by distributing independent linear-response $\mathbf{q}$-specific calculations of an atom-specific perturbation across the processors, each taking care of one or more $\mathbf{q}$ points. This parallelization is controlled using the input parameters \texttt{start\_q} and \texttt{last\_q} which specify the index of the starting and final $\mathbf{q}$ points from the list of all $\mathbf{q}$ points that have to be considered in the current pool of $\mathbf{q}$ points;
\item The $\mathbf{k}$~points parallelization, which is implemented by dividing
all processors into pools, each taking care of one or more $\mathbf{k}$ points;
\item The plane-wave parallelization, which is implemented by distributing real- and reciprocal-space grids across the processors.
\end{enumerate}
\begin{figure}[t]
\begin{center}
 \includegraphics[width=0.7\linewidth]{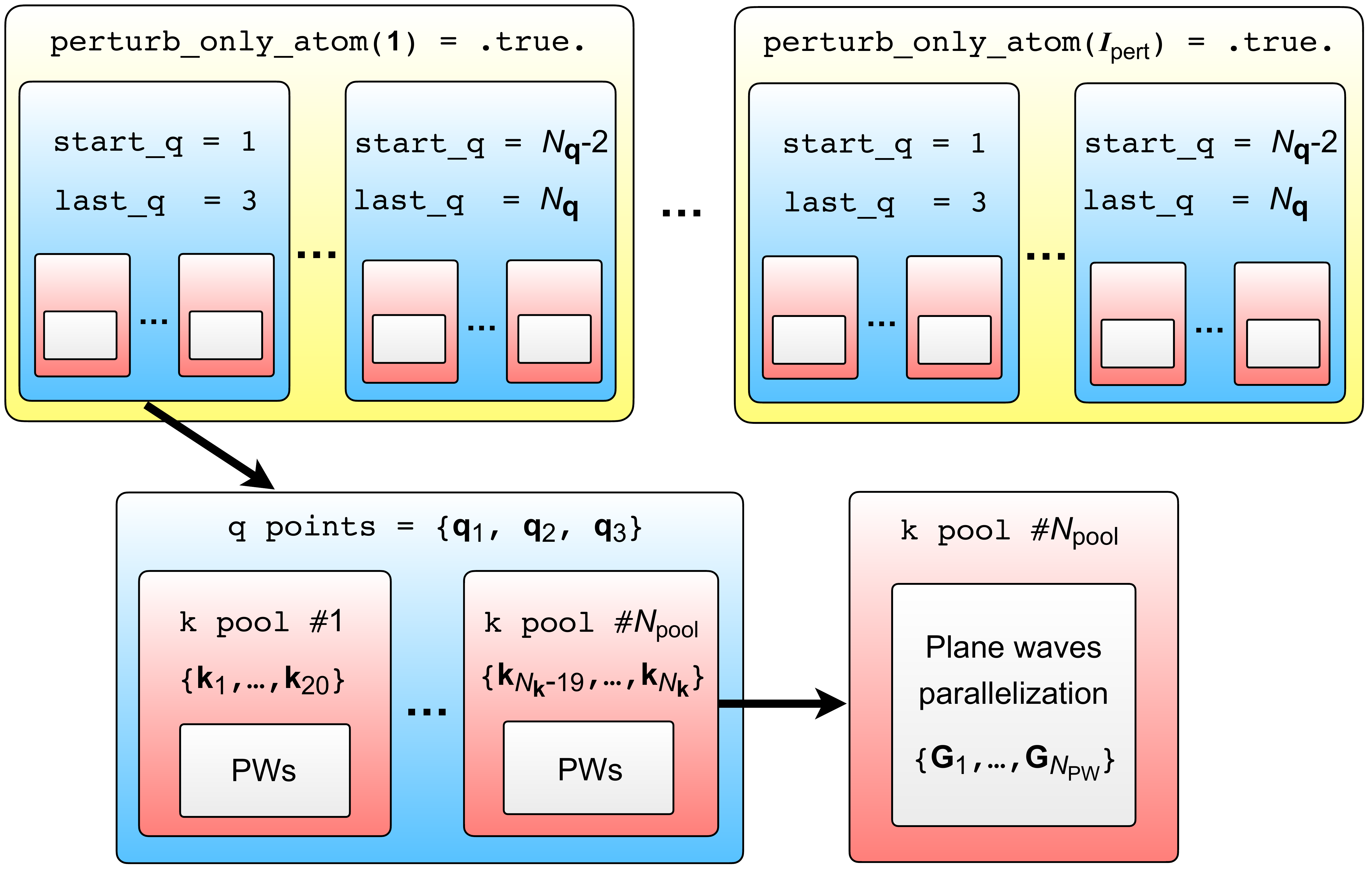}
 \caption{Schematic illustration of different parallelization levels of the \HP code. $N_\mathbf{k}$ and $N_\mathbf{q}$ are the number of $\mathbf{k}$ and $\mathbf{q}$ points, respectively, $N_\mathrm{pool}$ is the number of $\mathbf{k}$ points pools, $N_\mathrm{PW}$ is the number of plane waves (PWs) in the basis, and $I_\mathrm{pert}$ is the index of the perturbed Hubbard atom. The meaning of the keywords \texttt{perturb\_only\_atom}, \texttt{start\_q}, and \texttt{last\_q} is explained in Table~\ref{tab:Table_input_hp.x}. Note, the specific parallelization over the $\mathbf{k}$ and $\mathbf{q}$ points is shown just for demonstration purposes, but in practice the total number of $\mathbf{k}$ and $\mathbf{q}$ can be split in many different ways, i.e. different number of $\mathbf{k}$ point pools $N_\mathrm{pool}$ and different parallelization over $\mathbf{q}$ points (depending on the availability of computational resources).}
\label{fig1}
\end{center}
\end{figure}
Figure~\ref{fig1} shows the hierarchy of the parallelization levels of the \HP code. First, the calculation can be parallelized over the Hubbard atoms that must be perturbed (see \texttt{perturb\_only\_atom}); second, for each perturbed Hubbard atom the calculation can be parallelized over $\mathbf{q}$ points (see \texttt{start\_q} and \texttt{last\_q}); third, for each $\mathbf{q}$ point (or a subset of $\mathbf{q}$ points) the linear-response calculation can be parallelized over the $\mathbf{k}$ points (by choosing an appropriate number of $\mathbf{k}$ point pools $N_\mathrm{pool}$); and, last, within each $\mathbf{k}$ point pool all available CPUs are used to parallelize the calculation over plane waves ($\mathbf{G}$ points). If the system under study is quite small (say, a handful of atoms) and it can be run on a local workstation (with e.g. 8-16 cores), then it is convenient to skip the parallelization over perturbed Hubbard atoms and over $\mathbf{q}$ points, and use the parallelization over the $\mathbf{k}$ points (which can also be skipped) and use (only) the parallelization over plane waves. If the calculations are run on HPC clusters with many nodes, then it is highly recommended to use all aforementioned levels of parallelization in order to utilize the computational resources in the most effective way. Moreover, on HPCs it is recommended to avoid that the same compute node is split between $\mathbf{k}$ pools or groups of $\mathbf{q}$ points or perturbed atoms (due to the slow inter-node communications). Finally, we note that at present these levels of parallelization have to be chosen by the user manually.

\subsection{Testing of the code}
\label{sec:testing}

The \HP code implements DFPT which is complex from the programming point of view, and hence it is crucial to have a test suite to ensure that the new developments do not break existing functionalities of the code -- its availability facilitates maintenance of the code and ensures its long-term stability. As other components of \QE, the \HP code relies on the test suite that is based on the \texttt{testcode.py}~\cite{Spencer}: this provides the functionality to run tests automatically (nightly) and compare selected quantities (Hubbard parameters) parsed from the output files generated by \HP against reference values. The \HP code is run both in serial and in parallel, using various combinations of commonly used compilers [Intel Fortran compilers (ifort), GNU Fotran compilers (GFortran), etc.] and libraries (Intel MPI, Open MPI, etc.), which ensures that the code function correctly on various high-performance computer (HPC) architectures and in different environments. Whenever new features are added to the \HP code, the corresponding tests must be added to the test suite by the developers in order to guaranty the robustness of these features in the long term.

\section{Benchmarking}
\label{sec:benchmarking}

We now showcase how to use the \HP code for computing the Hubbard $U$ for the Fe($3d$) and Mn($3d$) states and Hubbard $V$ for Fe($3d$)--O($2p$) and Mn($3d$)--O($2p$) in Li$_x$Mn$_{1/2}$Fe$_{1/2}$PO$_4$ at $x=0$, $1/2$, and $1$. We recall that the validation of the DFPT implementation versus the finite-difference supercell approach of Refs.~\cite{Cococcioni:2005, Campo:2010} was already done in our previous works~\cite{Timrov:2018, Timrov:2021}. After computing $U$ and $V$ self-consistently, we present the results for this material obtained in the framework of DFT+$U$ and DFT+$U$+$V$.

\subsection{Technical details}
\label{sec:technical_details}

All calculations were performed using the plane-wave (PW) pseudopotential method as implemented in the \textsc{Quantum ESPRESSO} distribution~\cite{Giannozzi:2009, Giannozzi:2017, Giannozzi:2020}. We have used the xc functional constructed using spin-polarized GGA with the PBEsol prescription~\cite{Perdew:2008}. The PPs were taken from the SSSP library~v1.1 (efficiency)~\cite{prandini2018precision, MaterialsCloud}, which are either US or PAW: For manganese we have used \texttt{mn\_pbesol\_v1.5.uspp.F.UPF} from the GBRV v1.5 library~\cite{Garrity:2014}, for iron and oxygen \texttt{Fe.pbesol-spn-kjpaw\_psl.0.2.1.UPF} and \texttt{O.pbesol-n-kjpaw\_psl.0.1.UPF} from the Pslibrary v0.3.1~\cite{Kucukbenli:2014}, for phosphorus \texttt{P.pbesol-n-rrkjus\_psl.1.0.0.UPF} from the Pslibrary v1.0.0~\cite{DalCorso:2014}, and for lithium \texttt{li\_pbesol\_v1.4.uspp.F.UPF} from the GBRV v1.4 library~\cite{Garrity:2014}. To construct the Hubbard projector functions $\varphi^I_m(\mathbf{r})$ [see Eq.~\eqref{eq:Pm1m2}] we have used atomic orbitals which are orthogonalized using L\"owdin's method~\cite{Lowdin:1950, Mayer:2002}. Structural optimizations using DFT+$U$ and DFT+$U$+$V$ were performed using orthogonalized atomic orbitals as described in detail in Ref.~\cite{Timrov:2020b}. KS wavefunctions and potentials were expanded in PWs up to a kinetic-energy cutoff of 90 and 1080~Ry, respectively, and the BZ was sampled using the uniform $\Gamma$-centered $\mathbf{k}$ point mesh of size $5 \times 8 \times 9$, for structural optimization. The crystal structure was optimized using the Broyden-Fletcher-Goldfarb-Shanno (BFGS) algorithm~\cite{Fletcher:1987}, with a convergence threshold of $10^{-6}$~Ry for the total energy, $10^{-5}$~Ry/Bohr for forces, and 0.5~Kbar for pressure. For the metallic ground states we have used the Marzari-Vanderbilt smearing method~\cite{Marzari:1999} with a broadening parameter of 0.02~Ry. 

The DFPT calculations of Hubbard parameters were performed using the uniform $\Gamma$-centered $\mathbf{k}$ and $\mathbf{q}$ point meshes of size $3 \times 4 \times 5$ and $1 \times 2 \times 3$, respectively, which give an accuracy of 0.01~eV for the computed values of $U$ and $V$. These  $\mathbf{k}$ and $\mathbf{q}$ point meshes were determined by performing convergence tests as described in detail in Ref.~\cite{Timrov:2018}. The KS wavefunctions and potentials were expanded in PWs up to a kinetic-energy cutoff of 65 and 780~Ry, respectively, for calculation of Hubbard parameters. The linear-response KS equations of DFPT were solved using the conjugate-gradient algorithm~\cite{Payne:1992} and the mixing scheme of Ref.~\cite{Johnson:1988} for the response potential to speed up convergence.

Bulk Li was modeled at the DFT-PBEsol level using the \textit{bcc} unit cell with one Li atom at the origin. The optimized lattice parameter is 3.436~\AA. The BZ was sampled using the uniform $\Gamma$-centered $\mathbf{k}$ point mesh of size $10 \times 10 \times 10$, and we have used the Marzari-Vanderbilt smearing method~\cite{Marzari:1999} with a broadening parameter of 0.02~Ry. The KS wavefunctions and potentials were expanded in PWs up to a kinetic-energy cutoff of 65 and 780~Ry, respectively.

\begin{figure}[t]
  \centering
  \includegraphics[width=0.45\linewidth]{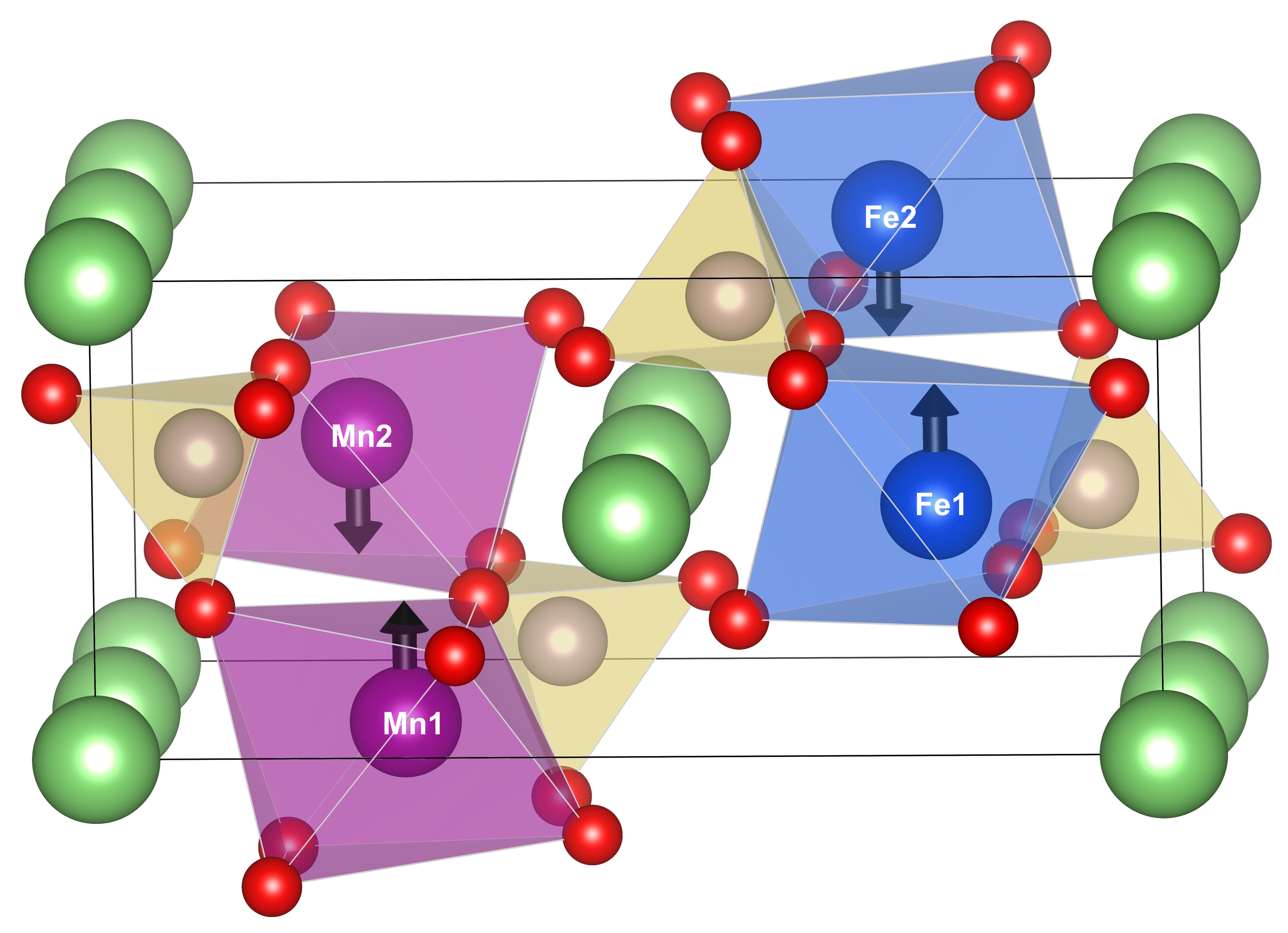}
   \caption{Crystal structure of the phospho-olivine LiMn$_{1/2}$Fe$_{1/2}$PO$_4$. Fe atoms are indicated in blue, Mn atoms in purple, O atoms in red, Li atoms in green, and P atoms in yellow. Black vertical arrows indicate the orientation of spin. Rendered using \textsc{VESTA}~\cite{Momma:2008}.}
\label{fig:crystal_structure}
\end{figure}

The phospho-olivine Li$_x$Mn$_{1/2}$Fe$_{1/2}$PO$_4$ has an orthorhombic crystal structure at $x=0$ and $x=1$ with a $Pnma$ space group~\cite{Muraliganth:2010}. The unit cell contains four formula units, i.e. 24 atoms in the case of $x=0$ and 28 atoms in the case of $x=1$. The crystal structure at $x=1$ is shown in Fig.~\ref{fig:crystal_structure}. The transition-metal (TM) atoms (labeled as $M$) are coordinated by six O atoms forming a $M$O$_6$ octahedron of which it occupies the center. The P atoms are instead at the center of PO$_4$ tetrahedra that they form with neighboring oxygens. The three-dimensional structure of the crystal can be understood as being based on a network of corner-sharing $M$O$_6$ octahedra further linked by ``interstitial'' PO$_4$ tetrahedra that act as structural reinforcer [avoiding excessive volume variations upon Li (de-)intercalation] and chemical stabilizers (useful to avoid oxygen escapes). Li ions reside within octahedral channels along the intermediate-length side of the cell. 
The phospho-olivines are known to show an antiferromagnetic behavior below their transition temperatures. In the previous study (Ref.~\cite{Cococcioni:2019}) it was shown that different antiferromagnetic arrangements of spins result in total energies that differ not more than by $\sim 20$~meV at the DFT+$U$+$V$ level of theory. Here we use the magnetic configuration that minimizes the total energy (labeled ``AF$_1$'' in Ref.~\cite{Cococcioni:2019}), and it is depicted in Fig.~\ref{fig:crystal_structure}. Finally, there are several configurations for arranging two Mn and two Fe atoms in the unit cell of Li$_x$Mn$_{1/2}$Fe$_{1/2}$PO$_4$. Our goal here is not to investigate all configurations but rather to choose one as a representative case for comparing results obtained using different levels of theory. To this end, we choose to arrange Mn and Fe atoms such that two Mn atoms are antiferromagnetically coupled to each other and same for Fe atoms, as shown in Fig.~\ref{fig:crystal_structure}.

The data used to produce the results of this work are available in the Materials Cloud Archive~\cite{MaterialsCloudArchive2022}.

\subsection{Ground-state calculation}
\label{sec:ground-state_calc}

In order to compute the Hubbard parameteres, independently from the specific functional adopted (DFT, DFT+$U$, or DFT+$U$+$V$), the first step is the ground-state calculation using the \PW code. Since we are interested in a self-consistent calculation of Hubbard parameters~\cite{Timrov:2021}, in \ref{sec:Sample_input_files} we show the input samples for DFT+$U$ (input sample 1) and DFT+$U$+$V$ (input sample 2) already with the converged values of $U$ and $V$. If one wants to start from the DFT ground-state, then the values of $U$ for Fe($3d$) and Mn($3d$) states have to be initialized to some small numbers (e.g. $10^{-10}$~eV) for the sake of activating the Hubbard-related machinery (in the DFT+$U$+$V$ case, there is no need to initialize $V$, instead initialize $U$ for O($2p$)). 

The input files for the \PW code contain the standard input parameters that will not be described here (the interested reader is invited to check the documentation of the \PW code~\cite{QuantumESPRESSO:website}). Instead, we focus on the initialization of the Hubbard-related input parameters. Since \QE v7.1, the input syntax for Hubbard-corrected DFT has changed to make it more user-friendly. More specifically, there is a new input card called ``HUBBARD'', where Hubbard-related data has to be specified. In particular, one has to specify the type of Hubbard projectors. Currently, in \QE there are two most popular types of Hubbard projectors, namely ``atomic'' that corresponds to the nonorthogonalzied atomic orbitals and ``ortho-atomic'' that corresponds to the L\"owdin orthogonalized atomic orbitals (see Sec.~\ref{sec:Hubbard_projectors}). Here we use the second option, thus in the input file we specify ``HUBBARD \{ortho-atomic\}''. Next, one has to specify the Hubbard manifolds, the values of the Hubbard parameters, and the indices of neighbors $I$ and $J$ between which $V$ is applied in the case of DFT+$U$+$V$. On the one hand, in the case of DFT+$U$ (see the input sample~1 in \ref{sec:Sample_input_files}) we specify that we want to apply the Hubbard $U$ correction to Fe1, Fe2, Mn1, and Mn2, so we write a letter \texttt{U} on each input line inside the ``HUBBARD'' card. For each of the Hubbard atomic types we specify the Hubbard manifold which is $3d$, hence we indicate it as \texttt{Fe1-3d}, \texttt{Fe2-3d}, \texttt{Mn1-3d}, and \texttt{Mn2-3d}. Finally, we specify the corresponding values of the Hubbard $U$ parameters on each input line: for Fe1($3d$) and Fe2($3d$) states we have exactly the same value of 4.97~eV, while for Mn1($3d$) and Mn2($3d$) we have 4.32~eV. These values were computed self-consistently using the \HP code (see below). It is important to remark that Fe1 and Fe2 are crystallographically equivalent, and same for Mn1 and Mn2, and they differ only by the orientation of spin (this is why we have defined sublattices with different indices). That is why the values of Hubbard $U$ parameters are the same for the same TM elements. On the other hand, in the case of DFT+$U$+$V$ we need to specify both Hubbard $U$ and $V$. The syntax for setting up Hubbard $U$ is the same as in the DFT+$U$ case. To specify Hubbard $V$, we need to indicate a letter \texttt{V} and then to indicate what is the couple of neighbors that we want to consider. For example, in the input sample~2 in \ref{sec:Sample_input_files} we specify the pair \texttt{Fe1-3d} and \texttt{O-2p}, and similarly for other nearest neighbors for each TM element. However, this is not all and we need to specify the indices $I$ and $J$ to say specifically to which atoms we are referring. Since \QE uses periodic boundary conditions, our real simulation cell is virtually replicated in all three Cartesian directions (positive and negative directions) and periodic replicas of atoms are generated (so we end up with a virtual $3 \times 3 \times 3$ supercell). It is important to stress that in practice we still work with the unit cell while the virtual $3 \times 3 \times 3$ supercell is only generated internally in the code just for the sake of determining the indices $I$ and $J$ of the neighboring atoms. This \textit{virtual} supercell should not be confused with any of the \textit{real} supercells that are used in the LR-cDFT approach to converge the computed Hubbard parameters. \textit{A priori} the user is not supposed to specify these indices, instead, these indices and the values of Hubbard parameters are obtained as an output of the \HP calculation. In the input sample~2 in \ref{sec:Sample_input_files} we can see that for each TM element we have specified six nearest neighbors because of the octahedral coordination; from our experience, setting six nearest neighbors is sufficient in the vast majority of cases. However, technically there is no restriction to include even further neighbors, but care must be taken in converging accurately the values of $V$.

In the case of magnetic insulators (which is the case here), the ground-state calculation must be performed using a two-step procedure. First step, we need to perform a self-consistent-field (SCF) DFT+$U$ or DFT+$U$+$V$ calculation as indicated in \ref{sec:Sample_input_files} by treating the system as a fake metal by using some smearing function (\texttt{occupations = 'smearing'}, \texttt{smearing = 'mv'}, and \texttt{degauss = 0.01} in the \texttt{system} namelist). This is needed technically in order to allow for fractional occupations in the spin-polarized calculation (\texttt{nspin = 2} in the \texttt{system} namelist) by setting some nonzero starting magnetization to each TM element (see \texttt{starting\_magnetization}). If we proceed directly to the \HP calculation immediately after this first step there will be a problem because the density of states at the Fermi level is very small and there will be a diverging ``metallic term'' (see Eq.~(79) in Ref.~\cite{Baroni:2001}). Therefore, we need to perform a second SCF calculation by restarting from the wavefunctions and spin charge density/potential of the first SCF calculation (by adding \texttt{startingwfc = 'file'} and \texttt{startingpot = 'file'} in the \texttt{electrons} namelist), by setting occupations to be fixed (\texttt{occupations = 'fixed'}), and by setting a total magnetization to be equal to the one determined in the first SCF calculation (\texttt{tot\_magnetization = 0.00} in this case in the \texttt{electrons} namelist). After this second SCF calculation we obtain a converged ground state for a magnetic insulator, which allows us to proceed to computing Hubbard parameters using DFPT as implemented in the \HP code.

\subsection{Hubbard parameters}
\label{sec:Hubbard_parameters}

In this section we discuss how to compute Hubbard parameters using the \HP code starting from the data generated from the ground-state calculation (see Sec.~\ref{sec:ground-state_calc}). The input file for the linear-response calculation is quite simple and is shown in \ref{sec:Sample_input_files} (see the input sample~3). First of all, one has to specify \texttt{prefix = 'olivine'} and \texttt{outdir='./'}, which are the prefix and output directory that must be exactly the same as in the input samples~1 or 2 (this is needed for reading the ground-state data). Second, the $\mathbf{q}$ point grid must be set: in this case we use the grid of size $1 \times 2 \times 3$ and we specify it as \texttt{nq1 = 1}, \texttt{nq2 = 2}, and \texttt{nq3 = 3}. The values of the Hubbard parameters computed using DFPT must be converged with respect to the size of the $\mathbf{q}$ point grid~\cite{Timrov:2018}; this is equivalent to converging Hubbard parameters with respect to the size of the (real) supercell when using the LR-cDFT approach of Ref.~\cite{Cococcioni:2005}. In principle, this should be enough in the majority of cases because the default values for other input parameters will be used. For the case of the phospho-olivine Li$_x$Mn$_{1/2}$Fe$_{1/2}$PO$_4$ we have changed somewhat other two input parameters compared to the default values. Namely, we set \texttt{conv\_thr\_chi = 1.0d-7}~eV$^{-1}$ which is the convergence threshold for the self-consistent response matrix $(\chi)_{IJ}$ [see Eq.~\eqref{chi}] during the iterative solution of the Sternheimer equation~\eqref{eq:LRKSeq_lp}, and \texttt{dist\_thr = 5.D-3}~Bohr which is the threshold for comparing inter-atomic distances when reconstructing the missing elements of the response susceptibility matrices $(\chi)_{IJ}$ and $(\chi_0)_{IJ}$ in the post-processing step.

It is useful to comment on the different options available to make the \HP code determine which Hubbard atoms must be perturbed. This is controlled by the input parameter \texttt{find\_atpert} (meaning ``find atoms to perturb''). The default value (which is also used in this work) is \texttt{find\_atpert}=1: it checks the ground-state atomic occupations, $\mathrm{Tr}[n^{I\sigma}_{m_1 m_2}]$, and compares them for different Hubbard atoms. If the differences between the atomic occupations of Hubbard atoms of the same type are smaller than the threshold \texttt{docc\_thr} (whose default value is $5 \times 10^{-5}$) then these Hubbard atoms are considered to be crystallographically equivalent (even if the magnetic moments have opposite sign). By applying this check the Hubbard atoms are classified according to their occupations and one atom per class is then perturbed. This option is the default one because it is general and it works well in most cases (regardless the number of symmetries that the system features). A second option is \texttt{find\_atpert}=2: the code perturbs one Hubbard atom per type. It is important to stress that when using this option, Hubbard atoms of the same type will be always treated as equivalent (for the purpose of calculating the Hubbard parameters) even if they are crystallographically inequivalent or show different atomic occupations. This could lead to inaccuracies if not used properly. This option is useful when the user wants to distinguish Hubbard atoms that might accidentally assume the same occupation (see previous option). Option \texttt{find\_atpert}=3 corresponds to determining Hubbard atom perturbation classes based on the symmetries of the system. This option, mostly useful when the system is highly symmetric, serves to distinguish Hubbard atoms that accidentally show the same atomic occupation. Finally, \texttt{find\_atpert}=4 corresponds to perturbing separately all the Hubbard atoms in the unit cell (i.e., assuming they are all inequivalent), which is obviously the most computationally expensive case. All these options of \texttt{find\_atpert} 
provide a significant flexibility in determining which Hubbard atoms should be perturbed in a system of interest. In general, however, it is recommended to start with the default option \texttt{find\_atpert}=1. By applying this procedure to LiMn$_{1/2}$Fe$_{1/2}$PO$_4$, 8 Hubbard atoms are perturbed in total within DFT+$U$+$V$: Fe1 (\#1), Mn1 (\#3), and 6 oxygen atoms (\#5, \#6, \#9, \#10, \#13, \#15) - see the numbering of atoms in the \texttt{ATOMIC\_POSITIONS} card in \ref{sec:Sample_input_files}. As can be seen, within DFT+$U$+$V$ when we compute the Hubbard parameters using DFPT we perturb also O atoms - this is needed in order to determine the inter-site Hubbard $V$ parameters. However, we do not report and do not use Hubbard $U$ for O atoms.

The resulting Hubbard parameters in the DFT+$U$ and DFT+$U$+$V$ frameworks are shown in Table~\ref{tab:hub_param}. It is interesting to observe how the values of Hubbard parameters change upon the lithiation of the material. In the DFT+$U$ case we can see that when going from $x=0$ to $x=1/2$ the $U$ value for Fe($3d$) states is only slightly increased (by 0.13~eV), while the $U$ value for Mn($3d$) states decreased significantly (by 1.66~eV). This means that the extra two electrons went to two Mn atoms and hence the corresponding $U$ values dropped substantially. When further going from $x=1/2$ to $x=1$ we see that the $U$ value for Fe has now decreased (by 0.17~eV), while $U$ for Mn has further decreased (by 0.25~eV). Such a nonmonotonic behavior of $U$ for Fe upon the lithiation is quite confusing, while for Mn we observe a systematic decrease with $x$. As will be shown in the following, this is a consequence of the fact that DFT+$U$ does not take into account the inter-site Hubbard interactions that are very important for materials with covalent bonding. 

\begin{table*}[h!]
\centering
\begin{tabular}{c|c|c|c|c|c|c}
\hline\hline
\parbox{1cm}{\centering Method} & \parbox{1cm}{\centering $x$}  & \parbox{1cm}{\centering HP} &  \parbox{2cm}{\centering Fe1} & \parbox{2cm}{\centering Fe2} & \parbox{2cm}{\centering Mn1} & \parbox{2cm}{\centering Mn2} \\ \hline
\multirow{3}{*}{DFT+$U$}
&           0           & $U$ &  5.01     & 5.01      & 6.23      & 6.23      \\
&           1/2         & $U$ &  5.14     & 5.14      & 4.57      & 4.57      \\
&           1           & $U$ &  4.97     & 4.97      & 4.32      & 4.32      \\ \hline
\multirow{6}{*}{DFT+$U$+$V$}
& \multirow{2}{*}{0}    & $U$ &  5.43     & 5.43      & 6.27      & 6.27      \\
&                       & $V$ & 0.60-1.12 & 0.60-1.12 & 0.55-1.05 & 0.55-1.05 \\ \cline{2-7}
& \multirow{2}{*}{1/2}  & $U$ &  5.44     & 5.44      & 4.81      & 4.81      \\
&                       & $V$ & 0.54-1.06 & 0.54-1.06 & 0.28-0.91 & 0.28-0.91 \\ \cline{2-7}
& \multirow{2}{*}{1}    & $U$ &  5.28     & 5.28      & 4.58      & 4.58      \\
&                       & $V$ & 0.41-0.89 & 0.41-0.89 & 0.42-0.80 & 0.42-0.80 \\
\hline\hline
\end{tabular}
\caption{Self-consistent Hubbard parameters (HP) in eV computed using DFPT in the DFT+$U$ and DFT+$U$+$V$ frameworks for Fe($3d$) and Mn($3d$) states in Li$_x$Mn$_{1/2}$Fe$_{1/2}$PO$_4$ for $x=0$, $1/2$, and $1$. This is the case study presented also in Ref.~\cite{Timrov:2022}.}
\label{tab:hub_param}
\end{table*}

In the DFT+$U$+$V$ case we take into account the inter-site Hubbard interactions hence we can see more clear trends in Table~\ref{tab:hub_param}. More specifically, when going from $x=0$ to $x=1/2$ the $U$ value for Fe($3d$) states stays essentially constant, while the $U$ value for Mn($3d$) states decreases (by 1.46~eV). This clearly shows that the extra two electrons went to two Mn atoms, while Fe atoms remain unaffected. When further going from $x=1/2$ to $x=1$ the $U$ value for Fe($3d$) states decreased (by 0.16~eV), but also the $U$ value for Mn($3d$) states decreased (by 0.23~eV, which is much smaller than when going from $x=0$ to $x=1/2$). This means that the extra two electrons now went mainly to Fe atoms. The relatively small change in $U$ for Mn($3d$) states seems to indicate that these states are very active chemically and hence they are very sensitive to changes in the chemical environment around them. In addition, it is important to remark that these are self-consistent values of Hubbard parameters, i.e. a structural optimization is performed for each concentration of $x$~\cite{Timrov:2021}. This might explain in part while the $U$ values for Mn($3d$) states still change when going from $x=1/2$ to $x=1$. The overall decrease in the $U$ values for Fe and Mn can be explained by the fact that the $3d$ manifolds of TM ions acquire an extra electron due to the insertion of Li; the more electrons there are in the Hubbard manifold, the weaker is the screened Coulomb interaction between them, and hence the $U$ value is smaller. Regarding the Hubbard $V$, in Table~\ref{tab:hub_param} we show the range of obtained values for different couples Fe($3d$)-O($2p$) and Mn($3d$)-O($2p$). The intersite Hubbard $V$ also decreases on average when going from $x=0$ to $x=1$. This latter observation is justified by the fact that the Li insertion leads to an increase of the cell volume and of the Mn--O bond lengths; the larger the bond lengths between two atoms, the smaller is the intersite Hubbard $V$ interaction.

\subsection{Accurate geometry and energetics from DFT+$U$+$V$}
\label{sec:results}

Using the self-consistent Hubbard parameters presented in Sec.~\ref{sec:Hubbard_parameters}, various ground-state properties of the phospho-olivine Li$_x$Mn$_{1/2}$Fe$_{1/2}$PO$_4$ can be computed, such as lattice parameters, electronic structure, atomic occupations, magnetic moments, Li intercalation voltages, and others. The reader is invited to check Ref.~\cite{Timrov:2022} where all these properties are discussed in great detail for several phospho-olivines including Li$_x$Mn$_{1/2}$Fe$_{1/2}$PO$_4$. Here, for the sake of demonstration purposes only, we highlight briefly the accuracy of predictions of lattice parameters and Li intercalation voltages.

\begin{table}[h!]
 \begin{center}
  \begin{tabular}{ccccc}
    \hline\hline
    LP          &   \parbox{1.5cm}{\centering DFT}   &  \parbox{1.5cm}{\centering DFT+$U$} & \parbox{1.5cm}{\centering DFT+$U$+$V$} &   \parbox{1.5cm}{\centering Expt.}   \\ \hline
    $a$         &  10.30  &   10.41  &    10.40    &  10.38    \\
    $b$         &   6.02  &    6.05  &     6.05    &   6.04    \\
    $c$         &   4.70  &    4.72  &     4.72    &   4.71    \\ 
    $V$         & 291.46  &  297.47  &   297.07    & 296.00    \\
    \hline\hline
  \end{tabular}    
 \end{center}
\caption{The equilibrium lattice parameters (LP) $a$, $b$, and $c$ (in \AA) and the volume $V$ (in \AA$^3$) of LiMn$_{1/2}$Fe$_{1/2}$PO$_4$ computed using DFT, DFT+$U$, and DFT+$U$+$V$ with self-consistent $U$ and $V$ (see Table~\ref{tab:hub_param}). The experimental data is from Ref.~\cite{Muraliganth:2010}. This is the case study presented also in Ref.~\cite{Timrov:2022}.}
\label{tab:lattice_parameters}
\end{table}

To the best of our knowledge, the experimental lattice parameters and volume for Li$_x$Mn$_{1/2}$Fe$_{1/2}$PO$_4$ are available only at $x=1$~\cite{Muraliganth:2010}. Hence in Table~\ref{tab:lattice_parameters} we show the optimized lattice parameters and the experimental one at $x=1$. It can be seen that DFT underestimates the lattice parameters and the cell volume while both DFT+$U$ and DFT+$U$+$V$ slightly overestimate them, with DFT+$U$+$V$ marginally overcoming DFT+$U$ in terms of accuracy. The effect of $V$ on the crystal structure of this material is very small though. This trend is consistent with the one we found for LiMnPO$_4$ in Ref.~\cite{Timrov:2021}. Hence, overall we find that DFT+$U$+$V$ gives the closest agreement with the experimental lattice parameters at $x=1$. In Ref.~\cite{Timrov:2022} the optimized lattice parameters are presented also for other values of $x$. 

\begin{figure*}[h!]
\centering
  \includegraphics[width=0.45\linewidth]{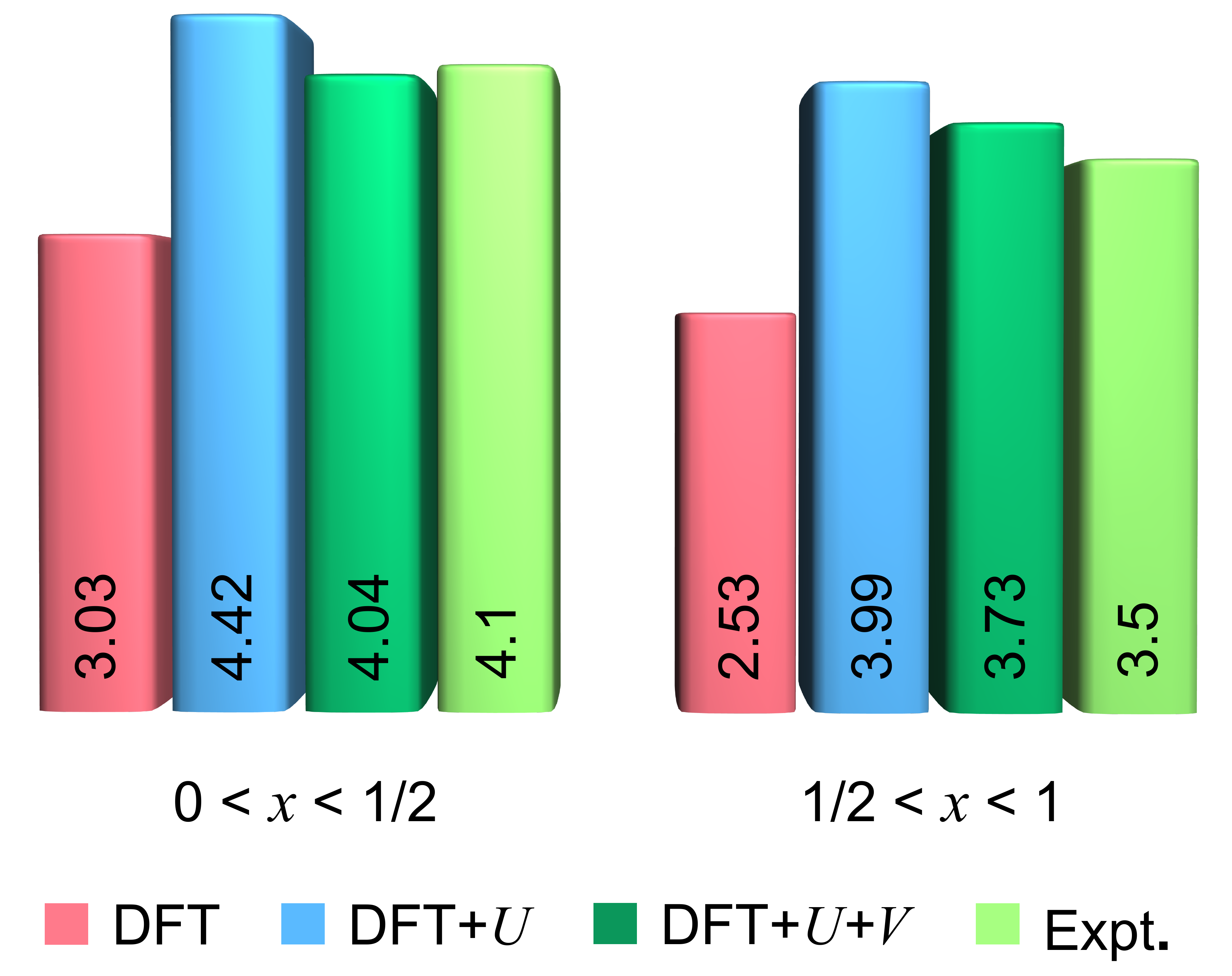}
   \caption{Voltages vs. Li/Li$^+$ (in V) for Li$_x$Mn$_{1/2}$Fe$_{1/2}$PO$_4$ for $0<x<1/2$ and $1/2<x<1$ computed using DFT, DFT+$U$, and DFT+$U$+$V$ with self-consistent $U$ and $V$ determined from first-principles (see Table~\ref{tab:hub_param}). The experimental data is from Ref.~\cite{Muraliganth:2010}. This is the case study presented also in Ref.~\cite{Timrov:2022}.}
\label{fig:voltages}
\end{figure*}

The Li intercalation voltages for Li$_x$Mn$_{1/2}$Fe$_{1/2}$PO$_4$ are shown in Fig.~\ref{fig:voltages}. A detailed discussion on how these voltages were computed can be found in Ref.~\cite{Timrov:2022}. Experimentally it is known that in Li-ion batteries with this material as a cathode there are two plateaus, with 4.1~V for $0<x<1/2$ and 3.5~V for $1/2<x<1$~\cite{Muraliganth:2010}. The former corresponds to the average voltage of Li$_x$MnPO$_4$, while the latter corresponds to the average voltage of Li$_x$FePO$_4$. Indeed, when intercalating Li and changing its concentration from $x=0$ to $x=1/2$, Mn ions are first to react while Fe ions remain unchanged, hence the voltage corresponds to the pristine material Li$_x$MnPO$_4$. When further increasing the concentration of Li from $x=1/2$ to $x=1$, now Fe ions react while Mn remain unchanged and hence the voltage corresponds to the pristine material Li$_x$FePO$_4$. As can be seen, DFT largely underestimates the voltages, while DFT+$U$ overestimates them. DFT+$U$+$V$ gives the best agreement with the experimental voltages, which means that applying the on-site $U$ correction alone is not sufficient and it is important to include also the inter-site $V$ correction to take properly into account the interactions of $3d$ electrons of TM ions with the $2p$ electrons of ligands. It is instructive to compare the accuracy of the DFT+$U$+$V$ voltages with the ones obtained when using HSE06. As reported in Ref.~\cite{Timrov:2022}, the HSE06 voltages (computed using the DFT+$U$+$V$ optimized geometry) are 4.34 and 4.03~V for $0 < x < 1/2$ and $1/2 < x < 1$, respectively. Therefore, we find that DFT+$U$+$V$ outperforms even HSE06 in terms of accuracy for predicting cathode voltages in this class of materials. Thus, DFT+$U$+$V$ provides the most accurate and reliable framework for predicting voltages in phospho-olivines~\cite{Cococcioni:2019, Timrov:2022}, and currently we are investigating the predictive accuracy of this approach for other types of cathode materials.

\section{Conclusions}
\label{sec:Conclusions}

We have presented the \HP code that implements DFPT for the calculation of the Hubbard parameters. DFPT allows to reduce significantly the computational costs and to improve the numerical accuracy of the Hubbard parameters by recasting the linear response to a localized perturbation into an array of monochromatic perturbations that can be calculated in the primitive cell independently of one another. Moreover, the calculation of empty electronic states is avoided~\cite{Baroni:2001} which greatly speeds up linear-response calculations of $U$ and $V$. 
The \HP code is one of the core components of the \QE distribution. It has multiple levels of parallelization which allows efficient usage of high-performance computers. Moreover, due to the high level of automation, \HP can be readily used for high-throughput calculations e.g. using \texttt{AiiDA}~\cite{Pizzi:2016, Huber:2020}.

The effectiveness of the code has been demonstrated by computing the Hubbard parameters self-consistently for the phospho-olivine Li$_x$Mn$_{1/2}$Fe$_{1/2}$PO$_4$ at $x=0$, $1/2$, and $1$. It has been shown that the Hubbard parameters change upon changes of $x$ which means that $U$ and $V$ should be recomputed at each Li concentration and not treated as global $x$-independent parameters. The predicted crystal geometry and intercalation voltages are in very good agreement with the experimental data, thus validating this statement. 

In the same spirit as the \QE project, \HP provides scientists worldwide with a well documented and open-source framework for implementing their ideas. It is in our best hope that \HP can benefit from the already well established users community of \QE for incorporating new ideas and keep growing in the future. The \HP code is hosted in a community accessible Git repository \cite{QuantumESPRESSO:Gitlab} and hence, apart from the releases of \QE~\cite{QuantumESPRESSO:website}, researchers who are willing to test the latest experimental implementations are welcome to do so and to contribute with their feedback.
Finally, the \HP code can be extended so as to employ various new features, in particular: (maximally-localized) Wannier functions as the projectors of the Hubbard manifold~\cite{Marzari:2012}; calculation of Hubbard parameters on top of meta-GGA functionals (e.g. SCAN~\cite{Sun:2015}); extension to multi-channel and noncollinear spin-polarized frameworks; ability to be run on novel GPU-enabled architectures, to name a few. These are some of the topics of future investigations of the \HP developers.

\section*{Acknowledgements}

We thank Paolo Giannozzi and Pietro Delugas for fruitful discussions. This research was supported by the Swiss National Science Foundation (SNSF), through grant 200021-179138, and its National Centre of Competence in Research (NCCR) MARVEL. M.C. acknowledges partial support from the EU-H2020 research and innovation programme under Grant Agreement No.~654360 within the framework of the NFFA Europe Transnational Access Activity. Computer time was provided by CSCS (Piz Daint) through project No.~s1073, and by the italian supercomputing center CINECA through the ISCRA project ``BatMat".

\appendix

\section{Sample input files}
\label{sec:Sample_input_files}

\noindent
{\bf Input sample 1: DFT+$U$ calculation using \texttt{pw.x}} 
\begin{verbatim}
&control
   calculation = 'scf', prefix = 'olivine', pseudo_dir = './', outdir = './'
/
&system
   ibrav = 8, celldm(1) = 19.667, celldm(2) = 0.582, celldm(3) = 0.454,
   nat = 28, ntyp = 7, nspin = 2, ecutwfc = 65.0, ecutrho = 780.0,
   occupations = 'smearing', smearing = 'mv', degauss = 0.01,
   starting_magnetization(1) = 0.5, starting_magnetization(2) = -0.5,
   starting_magnetization(3) = 0.5, starting_magnetization(4) = -0.5
/
&electrons
   conv_thr = 1.0d-10, mixing_mode = 'local-TF'
/
ATOMIC_SPECIES
Fe1  55.845  Fe.pbesol-spn-kjpaw_psl.0.2.1.UPF
Fe2  55.845  Fe.pbesol-spn-kjpaw_psl.0.2.1.UPF
Mn1  54.938  mn_pbesol_v1.5.uspp.F.UPF
Mn2  54.938  mn_pbesol_v1.5.uspp.F.UPF
O    15.999  O.pbesol-n-kjpaw_psl.0.1.UPF
P    30.974  P.pbesol-n-rrkjus_psl.1.0.0.UPF
Li   6.94    li_pbesol_v1.4.uspp.F.UPF
ATOMIC_POSITIONS {crystal}
Fe1   -0.0018269047        0.0000304069       -0.0003932705
Fe2   -0.0624820605        0.4999699836       -0.5003915084
Mn1    0.4987576776       -0.0000338875       -0.4510637530
Mn2    0.4369302234        0.5000334630       -0.9510599331
O     -0.1853704531       -0.0000413913       -0.2311704957
O      0.6251324004        0.4999517783       -0.7112993667
O      0.1210624075        0.5000390704       -0.7311671887
O      0.3105574643        0.0000452467       -0.2113000948
O      0.1689278301        0.0000409137       -0.7606906152
O      0.2595012636        0.5000440788       -0.1851692958
O     -0.2332375356        0.4999600895       -0.2606909453
O      0.6761898894       -0.0000458163       -0.6851718584
O      0.0510283716        0.7040582036       -0.1900775054
O      0.0510181642        0.2960368185       -0.1900725354
O      0.5580466406        0.7033668148       -0.2511808171
O      0.5580572495        0.2965326193       -0.2511588987
O     -0.1153361890       -0.2040586775       -0.6900811178
O     -0.1153274751        0.2039633923       -0.6900740032
O      0.3776426890       -0.2033681130       -0.7511796423
O      0.3776336406        0.2034650241       -0.7511611819
P     -0.1864198847       -0.0000468288       -0.5566953613
P      0.6287097290        0.4999490507       -0.3861250841
P      0.1221109842        0.5000464680       -0.0566921886
P      0.3069801138        0.0000492520       -0.8861258770
Li    -0.2786794632       -0.2501647928       -0.9729266705
Li     0.2143739303       -0.2498290476       -0.4729136212
Li     0.2144236723        0.2499658680       -0.4729252455
Li    -0.2787339024        0.2500407728       -0.9729169404
HUBBARD {ortho-atomic}
U  Fe1-3d  4.97
U  Fe2-3d  4.97
U  Mn1-3d  4.32
U  Mn2-3d  4.32
K_POINTS {automatic}
3 4 5 0 0 0
\end{verbatim}

\noindent
{\bf Input sample 2: DFT+$U$+$V$ calculation using \texttt{pw.x}} 
\begin{verbatim}
&control
    calculation = 'scf', prefix = 'olivine', pseudo_dir = './', outdir = './'
 /
 &system
    ibrav = 8, celldm(1) = 19.656, celldm(2) = 0.581, celldm(3) = 0.454,
    nat = 28, ntyp = 7, nspin = 2, ecutwfc = 65.0, ecutrho = 780.0,
    occupations = 'smearing', smearing = 'mv', degauss = 0.01,
    starting_magnetization(1) = 0.5, starting_magnetization(2) = -0.5,
    starting_magnetization(3) = 0.5, starting_magnetization(4) = -0.5
 /
 &electrons
    conv_thr =  1.0d-10, mixing_mode = 'local-TF'
 /
ATOMIC_SPECIES
Fe1  55.845  Fe.pbesol-spn-kjpaw_psl.0.2.1.UPF
Fe2  55.845  Fe.pbesol-spn-kjpaw_psl.0.2.1.UPF
Mn1  54.938  mn_pbesol_v1.5.uspp.F.UPF
Mn2  54.938  mn_pbesol_v1.5.uspp.F.UPF
O    15.999  O.pbesol-n-kjpaw_psl.0.1.UPF
P    30.974  P.pbesol-n-rrkjus_psl.1.0.0.UPF
Li   6.94    li_pbesol_v1.4.uspp.F.UPF
ATOMIC_POSITIONS {crystal}
Fe1   -0.0013002888        0.0000344681       -0.0011569560
Fe2   -0.0630088651        0.4999654781       -0.5011531240
Mn1    0.4991481712       -0.0000456327       -0.4508915833
Mn2    0.4365431830        0.5000453113       -0.9508863966
O     -0.1854167223       -0.0000457697       -0.2309578341
O      0.6251479692        0.4999513745       -0.7114114581
O      0.1211085328        0.5000433096       -0.7309509354
O      0.3105410503        0.0000450767       -0.2114131965
O      0.1689536157        0.0000428446       -0.7607925096
O      0.2594770427        0.5000509149       -0.1849993707
O     -0.2332639027        0.4999581183       -0.2607911995
O      0.6762141677       -0.0000526565       -0.6850067566
O      0.0509385486        0.7043511556       -0.1897007541
O      0.0509246526        0.2957600446       -0.1896986710
O      0.5579501365        0.7036161924       -0.2513817004
O      0.5579633292        0.2962483677       -0.2513649470
O     -0.1152470278       -0.2043524154       -0.6897058406
O     -0.1152343627        0.2042410671       -0.6897024045
O      0.3777385470       -0.2036180333       -0.7513800607
O      0.3777272053        0.2037488625       -0.7513678647
P     -0.1863309783       -0.0000548726       -0.5562820718
P      0.6286200092        0.4999421817       -0.3864261649
P      0.1220224059        0.5000536496       -0.0562763621
P      0.3070692847        0.0000558091       -0.8864285367
Li    -0.2786861788       -0.2501758028       -0.9729446405
Li     0.2143816342       -0.2498167962       -0.4729292757
Li     0.2144369823        0.2499681667       -0.4729426850
Li    -0.2787476689        0.2500403462       -0.9729317161
HUBBARD {ortho-atomic}
U  Fe1-3d  5.28
U  Fe2-3d  5.28
U  Mn1-3d  4.58
U  Mn2-3d  4.58
V  Fe1-3d  O-2p  1 321  0.89
V  Fe1-3d  O-2p  1  14  0.89
V  Fe1-3d  O-2p  1 401  0.78
V  Fe1-3d  O-2p  1   5  0.55
V  Fe1-3d  O-2p  1 410  0.42
V  Fe1-3d  O-2p  1 409  0.42
V  Fe2-3d  O-2p  2 465  0.89
V  Fe2-3d  O-2p  2  18  0.89
V  Fe2-3d  O-2p  2  11  0.78
V  Fe2-3d  O-2p  2   7  0.55
V  Fe2-3d  O-2p  2  14  0.42
V  Fe2-3d  O-2p  2  13  0.42
V  Mn1-3d  O-2p  3  16  0.80
V  Mn1-3d  O-2p  3 323  0.80
V  Mn1-3d  O-2p  3  12  0.76
V  Mn1-3d  O-2p  3  19  0.43
V  Mn1-3d  O-2p  3  20  0.43
V  Mn1-3d  O-2p  3   8  0.51
V  Mn2-3d  O-2p  4  20  0.80
V  Mn2-3d  O-2p  4 467  0.80
V  Mn2-3d  O-2p  4 374  0.76
V  Mn2-3d  O-2p  4 379  0.43
V  Mn2-3d  O-2p  4 380  0.43
V  Mn2-3d  O-2p  4   6  0.51
K_POINTS {automatic}
3 4 5 0 0 0
\end{verbatim}

\noindent
{\bf Input sample 3: calculation of $U$ (and $V$) using \texttt{hp.x}} 
\begin{verbatim}
&inputhp
   prefix = 'olivine', outdir='./',
   nq1 = 1, nq2 = 2, nq3 = 3,
   conv_thr_chi = 1.0d-7,
   dist_thr = 5.D-3
/
\end{verbatim}


\section{Input Variables}
\label{sec:Input_variables}

\begin{longtable}[c]{|>{\centering}m{3.5cm} |>{\centering}m{1.5cm} |>{}m{7.5cm}|}
    \hline
    Variable name  & Default & \hspace{2.5cm} Description \\ 
    \hline
    \endfirsthead
    \hline
    Variable name  & Default & \hspace{2.5cm} Description \\ 
    \hline
    \endhead
    \hline
    \multicolumn{3}{|c|}{Continuation on the next page...}\\
    \hline
    \endfoot
    \hline
    \caption{Input variables for the \HP code (\texttt{hp.x} executable), that have to be specified in the namelist \texttt{inputhp}. Here, only the most relevant input parameters are discussed; the complete list of all possible variables can be found in the documentation of the \HP code, which can be found in the folder /HP/Doc residing in the \QE tree.} \\
    \endlastfoot
    \multicolumn{3}{|c|}{\textbf{Basic keywords}}\\
    \hline
    \texttt{prefix} & '\texttt{pwscf}' & {\footnotesize 
    Prefix which is prepended to input/output filenames; must be the same used in the calculation of unperturbed system.} \\ 
    \texttt{outdir} & './' & {\footnotesize
    Path to the working directory containing temporary files (wavefunctions, spin charge density, occupation matrix, XML file with the system's data, etc., which are generated by a ground-state \texttt{pw.x} run).} \\ 
    \texttt{find\_atpert} & 1 & {\footnotesize
    Method for searching of atoms which must be perturbed. Possible values: 1 - find how many inequivalent Hubbard atoms there are by analyzing the trace of unperturbed occupation matrices, $\mathrm{Tr}[n^{I\sigma}_{m_1 m_2}]$; 2 - find how many Hubbard atoms to perturb based on how many different Hubbard atomic types there are. Note: atoms which have the same type but which are inequivalent by symmetry or which have different occupations will not be distinguished in this case; 3 - find how many inequivalent Hubbard atoms there are using symmetry. Atoms which have the same type but are not equivalent by symmetry will be distinguished in this case; 4 - perturb all Hubbard atoms (this is the most expensive option).} \\ 
    \texttt{docc\_thr} & $5 \times 10^{-5}$ (unitless) & {\footnotesize
    Threshold for the comparison of the traces of unperturbed occupation matrices $\mathrm{Tr}[n^{I\sigma}_{m_1 m_2}]$ of different Hubbard atoms, which is needed for the selection of atoms which must be perturbed. Can be used only when \texttt{find\_atpert} = 1.} \\
    \texttt{nq1, nq2, nq3} & 1, 1, 1 & \footnotesize{Size of the $\mathbf{q}$ point grid. Each $\mathbf{q}$ point corresponds to a monochromatic perturbation of electronic occupations in the Hubbard manifold of Hubbard atoms.} \\ 
    \texttt{skip\_equivalence\_q} & \texttt{.false.} & {\footnotesize
    If \texttt{.true.} then \texttt{hp.x} will skip the equivalence analysis of $\mathbf{q}$ points, and thus the full grid of $\mathbf{q}$ points will be used. Otherwise, the symmetry is used to determine equivalent $\mathbf{q}$ points (star of $\mathbf{q}$), and then perform calculations only for inequivalent $\mathbf{q}$ points.} \\
    \texttt{dist\_thr} & $6 \times 10^{-4}$ (Bohr) & {\footnotesize
    Threshold for comparing inter-atomic distances when reconstructing the missing elements of the response susceptibility matrices in the post-processing step.} \\
    \texttt{iverbosity} & 1 & {\footnotesize Verbosity level, i.e. the amount of information printed in the output file of the \texttt{hp.x} run. Possible values: 1 - minimal output; 2 - as 1 plus symmetry matrices, final response matrices $\chi_0$ and $\chi$ [see Eq.~\eqref{chi}] and their inverse matrices $\chi_0^{-1}$ and $\chi^{-1}$, and the matrix of Hubbard parameters ($U$ on the diagonal, $V$ on the off-diagonal); 3 - as 2 plus various details about the non-self-consistent calculation at $\mathbf{k}$ and $\mathbf{k+q}$ points; 4 - as 3 plus response occupation matrices $\Delta^{s'}_\mathbf{q} \bar{n}^{s \, \sigma}_{m_1 m_2}$ [see Eq.~\eqref{eq:occ_matrix_response_lp_2}].} \\ 
    \hline
    \multicolumn{3}{|c|}{\textbf{Keywords controlling the convergence}}\\
    \hline
    \texttt{conv\_thr\_chi} & $10^{-5}$ (eV$^{-1}$) & {\footnotesize 
    Convergence threshold for the self-consistent response matrix $\chi$ [see Eq.~\eqref{chi}] during the iterative solution of the Sternheimer equations~\eqref{eq:LRKSeq_lp}.} \\
    \texttt{thresh\_init} & $10^{-14}$ & {\footnotesize 
    Initial threshold for the solution of the Sternheimer equations (first iteration). Needed to converge the bare (non-interacting) response matrix $\chi_0$ [see Eq.~\eqref{chi}]. The specified value will be multiplied by the number of electrons in the system (i.e. it is an extensive quantity).} \\
    \texttt{ethr\_nscf} & $10^{-11}$ \par (Ry) & {\footnotesize 
    Threshold for the convergence of KS eigenvalues during the iterative diagonalization of the Hamiltonian in the non-self-consistent calculation at $\mathbf{k}$ and $\mathbf{k+q}$ points. Note, this quantity is not extensive.} \\
    \texttt{niter\_max} & 100 & {\footnotesize 
    Maximum number of iterations for the self-consistent iterative solution of the Sternheimer equations~\eqref{eq:LRKSeq_lp}.} \\
    \texttt{alpha\_mix(i)} & 0.3 & {\footnotesize 
    Mixing parameter (for the \texttt{i}-th iteration, \texttt{i} runs from 1 to \texttt{niter\_max}) for updating the response Hxc potential $\Delta^{s'}_\mathbf{q} \hat{\bar{V}}_{\mathrm{Hxc},\sigma}$ using the modified Broyden
    method~\cite{Johnson:1988}.} \\
    \texttt{nmix} & 4 & {\footnotesize 
    Number of iterations used in the mixing of the response Hxc potential $\Delta^{s'}_\mathbf{q} \hat{\bar{V}}_{\mathrm{Hxc},\sigma}$
    using the modified Broyden method~\cite{Johnson:1988}.} \\
    \hline
    \multicolumn{3}{|c|}{\textbf{Keywords for the parallelization of the calculations (optional)}}\\
    \hline
    \texttt{perturb\_only\_atom(i)} & \texttt{.false.} & {\footnotesize
    If \texttt{perturb\_only\_atom(i)=.true.} then only the \texttt{i}-th atom (not the atomic type) will be perturbed and considered in the run. This variable is useful when one wants to parallelize the whole calculation over perturbed Hubbard atoms (see Sec.~\ref{sec:parallelization}).} \\ 
    \texttt{start\_q} & 1 & {\footnotesize
    This keyword is used for the parallelization of the calculation over $\mathbf{q}$ points [see Sec.~\ref{sec:parallelization})] for a fixed perturbed atom (see \texttt{perturb\_only\_atom}). \texttt{start\_q} specifies the $\mathbf{q}$ point starting from which the calculations will be performed; see also \texttt{last\_q}.} \\ 
    \texttt{last\_q} & $N_\mathbf{q}$ & {\footnotesize
    This keyword is used for the parallelization of the calculation over $\mathbf{q}$ points [see Sec.~\ref{sec:parallelization})] for a fixed perturbed atom (see \texttt{perturb\_only\_atom}). \texttt{last\_q} specifies the $\mathbf{q}$ point up to which the calculations will be performed; see also \texttt{start\_q}.} \\ 
    \texttt{sum\_pertq} & \texttt{.false.} & {\footnotesize
    If it is set to \texttt{.true.} then \texttt{hp.x} will collect pieces of the response occupation matrices $\Delta^{s'}_\mathbf{q} \bar{n}^{s \, \sigma}_{m_1 m_2}$ [see Eq.~\eqref{eq:occ_matrix_response_lp_2}] for all $\mathbf{q}$ points and compute the sum of them including the respective phase factors and using the normalization $1/N_\mathbf{q}$, according to Eq.~\eqref{eq:occ_matrix_response_lp_1}. This variable should be used only when \texttt{start\_q}, \texttt{last\_q} and \texttt{perturb\_only\_atom} are used; otherwise, \texttt{hp.x} will automatically compute the sum using Eq.~\eqref{eq:occ_matrix_response_lp_1}.} \\
    \texttt{compute\_hp} & \texttt{.false.} & {\footnotesize
    This keyword is used to perform post-processing calculation of the Hubbard parameters.
    If it is set to \texttt{.true.}, \texttt{hp.x} will not perform linear-response calculations; instead, it will assume that selected columns of the $\chi_0$ and $\chi$ matrices were already computed in previous runs [each column corresponds to the response of occupations on all atoms to a perturbation of a specific Hubbard atom, see Eq.~\eqref{chi}]. The \texttt{hp.x} code will look for the files \texttt{prefix.chi.i.dat} (\texttt{i} runs over perturbed Hubbard atoms $I$) that must be stored in \texttt{outdir/HP/}. This keyword must be set to \texttt{.true.} when the calculation was parallelized over perturbations (or when the post-processing step must be re-run). \texttt{outdir} and \texttt{prefix} are defined in \ref{sec:Sample_input_files}.} \\
 \label{tab:Table_input_hp.x}
\end{longtable}


\begin{thebibliography}{100}
\bibitem{Timrov:Note:2020:nij}
{It is important to note that the response of the non-diagonal (with respect to
  the atomic sites) components of the occupation matrix in
  Eq.~\eqref{eq:occ_matrix_0} is not needed, because these terms are not
  present in the definition of the response matrices $(\chi_0)_{IJ}$ and
  $(\chi)_{IJ}$ (see Eq.~\eqref{chi}).}
\end{thebibliography}

\end{document}